# Pourquoi modéliser ?
# Le visuel comme reflet des perspectives intellectuelles en histoire médiévale[1]

**Plan de l'article**
1. Établir une typologie : visuels, graphiques, modèles
    1.1. Fonctions des « images » en médiévistique
    1.2 Fréquence et répartition du visuel dans un corpus scientifique (SHMESP)
    1.3. L'absence du visuel : causes et conséquences supposées
2. Du visuel au modèle : passage(s)
    2.1. L'historien, le corpus, la ruine
    2.2. Visualiser la production diplomatique européenne ?
    2.3. Le cas des édifices dits « romans »
3. Représenter le texte
    3.1. Mots, formules, formulaires
    3.2. Distinguer les « auteurs »
    3.3. Sémantique historique et visualisation
Éléments conclusifs

**Résumé**
L'article interroge l'importance des représentations graphiques dans les sciences sociales et particulièrement en histoire (médiévale), en partant d'une réflexion d'Étienne-Jules Marey, un physiologiste, pionnier de la photographie et du cinéma du XIX$^e$ siècle. Ce dernier estimait que le visuel devait remplacer le langage dans de nombreux domaines. De fait, le XX$^e$ et le début du XXI$^e$ siècle ont vu une multiplication exponentielle des supports visuels, en particulier avec l'avènement du numérique. Cependant, cette « révolution graphique » n'a pas touché toutes les disciplines de manière égale. Des différences importantes subsistent en fonction des domaines scientifiques, comme l'astrophysique, l'anthropologie, la chimie et l'histoire médiévale, malgré leur engagement commun à décrire des processus dynamiques et des changements d'état. Or, si les historiens ont déjà numérisé une grande partie du patrimoine culturel de l'Antiquité jusqu'aux X$^e$-XIII$^e$ siècles, l'exploration de ce corpus à l'aide de visualisations reste limitée. Il existe donc un potentiel inexploité dans ce domaine.

L'article vise tout d'abord à esquisser une typologie et une quantification des rôles passés et potentiels des représentations visuelles en histoire médiévale. Il examine deux approches intellectuelles distinctes : 1. l'utilisation des visuels pour soutenir un discours scientifique (majoritaire) et 2. la construction d'un discours historique fondé sur des observations réalisées à partir de figures visuelles ayant pour objectif de modéliser des phénomènes invisibles à l'œil nu. L'auteur examine ainsi l'utilisation des « images » en médiévistique, en se concentrant sur

---

[1] Une première version de cet article a été donnée sous la forme d'une communication dans le séminaire I.V.I. (*Idée, Vérité, Image – Eidos, alêtheia, eidôlon*, CRHI, CEPAM, MSHS-Sud Est), coorganisé en octobre 2016 par Elsa Grasso et Arnaud Zucker, à l'Université de Nice-Sophia Antipolis. Nous les remercions vivement tous deux, ainsi que Michel Lauwers, Rosa-Maria Dessi et Philippe Jansen, pour leurs remarques bienveillantes à l'occasion de cette intervention. Il a été par la suite présenté sous une forme remaniée, lors du colloque coorganisé par Hervé Duschêne et Daniel Russo (Centre Georges Chevrier, Université de Bourgogne), *Figures et Objets. Histoire, histoire de l'art, archéologie. Un moment intellectuel à travers ses figures et ses objets d'étude*. À cette occasion, nous remercions chaleureusement les organisateurs, ainsi qu'Annamaria Ducci et Pierre-Alain Mariaux, pour leurs précieuses remarques. Enfin, la version actuelle de l'article a donné lieu à plusieurs critiques et commentaires, dont le texte a grandement profité : toute notre gratitude va ainsi à Eliana Magnani, Didier Méhu, Joseph Morsel et Daniel Russo. <span style="color:red">Achevé en 2017, l'article devait initialement s'intégrer dans un volume collectif qui n'a pas vu le jour. Nous le publions tel quel en 2023, via HAL SHS, dans sa version de 2017 – y compris en ce qui concerne les références, arrêtées à cette date.</span>

les volumes annuels de la Société des historiens médiévistes de l'enseignement supérieur (SHMESP), jusqu'en 2006. Deux autres parties du texte se penchent sur les formes encore rares de représentations visuelles en histoire médiévale, notamment celles à « vocation heuristique », utilisant des objets iconographiques, des parchemins, des édifices et des textes numérisés. L'article suggère diverses techniques de visualisation, telles que l'analyse de réseaux, la création de « stemmas 2.0 » et des chronologies interactives, qui pourraient être bénéfiques pour la discipline. Ces méthodes pourraient potentiellement faire évoluer en profondeur notre compréhension des sociétés anciennes, en montrant les relations dynamiques entre différents aspects de ces sociétés. L'une des avancées les plus importantes attendues de ces méthodes visuelles est une meilleure compréhension des modèles de l'essor de l'Europe médiévale, qui fut variable d'une région à l'autre.

L'hypothèse soutenue est que la rareté des graphiques à vocation heuristique en histoire médiévale découle des rapports entretenus avec les documents anciens et de la méthode historique basée sur la narration et l'exemplarité. L'article interroge ainsi l'intérêt de la « modélisation visuelle » en histoire médiévale et souligne les défis liés à l'adoption généralisée de cette approche dans les sciences humaines et sociales. Enfin, le texte invite à réfléchir sur la nature et le fonctionnement des dispositifs visuels à vocation heuristique, en comparant les « images » médiévales et les visuels scientifiques contemporains. Dans les deux cas, il s'agit de matérialiser l'invisible pour montrer quelque chose qui existe au-delà du visuel. L'auteur suggère que cette manière d'aborder les visuels pourrait jouer un rôle croissant dans les décennies à venir, notamment dans le domaine des sciences des données.

**Abstract**
This article examines the importance of graphic representations in the social sciences, and particularly in (medieval) history, taking as its starting point a reflection by Étienne-Jules Marey, a physiologist and pioneer of 19th-century photography and cinema. Marey believed that the visual should replace language in many fields. Indeed, the twentieth and early twenty-first centuries saw an exponential multiplication of visual media, particularly with the advent of digital technology. However, this "graphics revolution" has not affected all disciplines equally. Significant differences remain between scientific fields such as astrophysics, anthropology, chemistry and medieval history, despite their shared commitment to describing dynamic processes and changes of state. Yet, while historians have already digitized a large part of the cultural heritage from Antiquity to the 10th-13th centuries, exploration of this corpus using visualizations remains limited. There is therefore untapped potential in this field.

This article begins by outlining a typology and quantification of the past and potential roles of visual representations in medieval history. It examines two distinct intellectual approaches: 1. the use of visuals to support a scientific discourse (majority) and 2. the construction of a historical discourse based on observations made from visual figures with the aim of modeling phenomena invisible to the naked eye. The author thus examines the use of "images" in medievalism, focusing on the annual volumes of the Société des historiens médiévistes de l'enseignement supérieur (SHMESP), up to 2006. Two other parts of the text look at the still-rare forms of visual representation in medieval history, particularly those with a "heuristic vocation", using iconographic objects, parchments, buildings and digitized texts. The article suggests various visualization techniques, such as network analysis, the creation of "stemmas 2.0" and interactive chronologies, which could benefit the discipline. These methods could potentially profoundly change our understanding of ancient societies, by showing the dynamic relationships between different aspects of these societies. One of the most important advances expected from these visual methods is a better understanding of the patterns of development in medieval Europe, which varied from region to region.

The hypothesis is that the scarcity of heuristic graphics in medieval history stems from the relationship with ancient documents and the historical method based on narration and exemplarity. The article thus questions the value of "visual modelling" in medieval history, and highlights the challenges associated with the widespread adoption of this approach in the humanities and social sciences. Finally, the text invites us to reflect on the nature and functioning of heuristic visual devices, by comparing medieval "images" and contemporary scientific visuals. In both cases, the point is to materialize the invisible in order to show something that exists beyond the visual. The author suggests that this way of approaching visuals could play a growing role in the decades to come, particularly in the field of data science.

En 1878, le professeur au Collège de France Étienne-Jules Marey [1830-1904], médecin, physiologiste, pionnier de la photographie et du cinéma, fait paraître son ouvrage *La méthode graphique dans les sciences expérimentales et principalement en physiologie et en médecine*[2]. Il y exprime sa pensée sur l'intérêt du visuel dans la compréhension des lois géologiques, biologiques mais aussi sociales : « Il n'est pas douteux que l'expression graphique ne se substitue bientôt à toute autre, chaque fois qu'il s'agira de définir un mouvement ou un changement d'état, en un mot un phénomène quelconque. Né avant la science et n'étant pas fait pour elle, le langage est souvent impropre à exprimer des mesures exactes, des rapports bien définis »[3]. L'expression graphique, pour reprendre celle de l'auteur, serait donc supérieure au langage, car elle permettrait une formalisation plus efficace de la pensée humaine et des informations que la cognition doit traiter. Dans une optique tout à fait darwinienne, la visualisation des données aurait donc été appelée, pour Marey, à remplacer le langage dans de nombreux domaines, puisqu'elle était, si l'on peut dire, plus adaptée « à son milieu » que ce dernier[4].

De fait, l'histoire du XX[e] et du début du XXI[e] siècle semble en partie avoir donné raison à l'auteur et à sa prédiction : les supports visuels se sont multipliés de façon exponentielle tout au long de cette période, mue par l'apparition d'un nouveau système-technique dont le numérique est la dernière forme en date[5]. Mais il est tout aussi évident que cette « révolution graphique »[6], objet d'étude du tournant iconique[7], n'a pas touché équitablement tous les secteurs de la société, y compris de la science[8]. Le visuel apparaît pourtant comme une voie royale de mise en forme et de mise en ordre des données scientifiques[9]. Or, il semble que la représentation graphique ne joue pas le même rôle aujourd'hui en astrophysique qu'en anthropologie, en chimie qu'en histoire médiévale – alors que toutes ces disciplines se chargent pourtant de décrire des « changements d'états », « des mouvements » (pour reprendre le vocabulaire de Marey), ou plutôt des « processus ». Au-delà des différences liées aux données analysées, pourquoi et comment subsistent de telles disparités ? Est-il seulement souhaitable de dépasser ce clivage ? Dès 1954, Claude Lévi-Strauss, au sein de l'article qu'il consacrait alors aux *Mathématiques de l'homme*, s'interrogeait : « Faut-il en conclure qu'entre les sciences exactes et naturelles, d'une part, les sciences humaines et sociales, de l'autre, la différence est si profonde, si irréductible, qu'on doit perdre tout espoir d'étendre jamais aux secondes les méthodes rigoureuses qui ont assuré le triomphe des premières ? »[10]. Les procédés graphiques sont, nous semble-t-il, au cœur de cette question ainsi que du clivage, essentiellement historique, disciplinaire, interprétatif, et non pas ontologique ou herméneutique, entre sciences dites dures et sciences sociales.

---

[2] É-J. MAREY, *La méthode graphique dans les sciences expérimentales et principalement en physiologie et en médecine*, Paris, 1885 ; M. BRAUN, *Picturing Time. The Work of Etienne-Jules Marey (1830-1904)*, Chicago, 1992 ; F. DAGOGNET, *Étienne-Jules Marey. La passion de la trace*, Paris, 1997 ; L. MANNONI, *Étienne-Jules Marey, la mémoire de l'œil*, Milan, 1999.

[3] É-J. MAREY, *La méthode graphique dans les sciences expérimentales et principalement en physiologie et en médecine*, op.cit., p. iii. Le passage est cité en anglais dans L. DASTON et P. GALISON, « The Image of Objectivity », in *Representations*, 40 (1992), p. 81-128.

[4] Sur l'historicité de la notion d'objectivité en science, en particulier à travers les dispositifs visuels, voir L. DASTON et P. GALISON, *Objectivity*, New York, 2007 ; C. ALLAMEL-RAFFIN, « La complexité des images scientifiques. Ce que la sémiotique de l'image nous apprend sur l'objectivité scientifique », in *Communication et langages*, 149 (2006), p. 97-111. Sur le développement de ces mêmes dispositifs, on pourra consulter : L. DASTON et E. LUNBECK (éd.), *Histories of Scientific Observation*, Chicago, 2011.

[5] Sur l'impact de la reproduction mécanisée dans le domaine esthétique, voir W. BENJAMIN, *L'œuvre d'art à l'époque de sa reproductibilité technique*, in *Œuvres III*, Paris, 2000 (1935 pour la rédaction). Plus récemment, dans le champ des *Visual Studies* : E. ALLOA (éd.), *Penser l'image*, Dijon, 2010, en particulier H. BREDEKAMP, « La « main pensante ». L'image dans les sciences », p. 177-209. Voir de même les travaux de Martin Kemp sur le visuel dans la revue *Nature* : M. KEMP, *Bilderwissen. Die Anschaulichkeit naturwissenschaftlicher Phänomene*, Cologne, 2003.

Cet article a donc comme objectif premier d'esquisser une typologie, une quantification ainsi qu'une définition des rôles à la fois passés et potentiels des « représentations visuelles » en histoire médiévale, branche certes spécifique de l'Histoire mais aussi relativement représentative des pratiques historiennes en général[11]. Dans le cadre du présent volume, il s'agit ainsi de *comprendre un moment intellectuel à travers ses figures*, ou plutôt *deux moments intellectuels* qui correspondent à différences manières d'envisager le rôle des figures scientifiques en tant que paradigmes[12]. L'un, certes riche mais classique, qui s'ancre dans l'héritage intellectuel issu du milieu du XIX[e] siècle : il consiste à employer le visuel afin d'appuyer un discours scientifique, à titre exemplaire. L'autre, qui trouve aussi son origine dans la période 1880-1930, mais qui n'a pas connu le même succès jusqu'ici, présente de belles perspectives pour l'avenir. Dans ce deuxième cas, il s'agit de construire le discours historien en s'appuyant sur les figures, qui ont alors pour fonction de modéliser des phénomènes, et qui sont placées en amont de la procédure analytique.

---

[6] J. GOODY, *La raison graphique. La domestication de la pensée sauvage*, Paris, 1979.

[7] G. BOEHM, « Die Wiederkehr der Bilder », in *Was ist ein Bild?*, Munich, 1994, p. 11-38 ; W.J.T. MITCHELL, *Que veulent les images ? Une critique de la culture visuelle*, Dijon, 2014.

[8] Un peu en retrait en France, les études sur le visuel scientifique se multiplient dans les mondes anglo-saxon et allemand. Outre les ouvrages de Lorraine Daston déjà mentionnés, voir M. SICARD, *La fabrique du regard. Images de science et appareils de vision (XV$^e$-XX$^e$ siècle)*, Paris, 1998 ; L. PAUWELS (éd.), *Visual Cultures of Science: Rethinking Representational Practices in Knowledge Building and Science Communication*, Lebanon, 2006 ; R. MASON, *Reading Scientific Images. The Iconography of Evolution*, Cape Town, 2006 ; C. BIGG, « Les études visuelles des sciences : regards croisés sur les images scientifiques », in *Histoire de l'art*, 70 (juillet 2012), p. 95-101, qui résume utilement la bibliographie essentielle jusqu'à cette date ; A. CARUSI, A. SISSEL HOEL, T. WEBMOOR, S. WOOLGAR (dir.), *Visualization in the Age of Computerization*, New York, 2014 ; C. COOPMANS, J. VERTESI, M. LYNCH et S. WOOLGAR (éd.), *Representation in Scientific Practice Revisited*, Cambridge, 2014 ; K. HENTSCHEL, *Visual Cultures in Science and Technology: A Comparative History*, Oxford, 2014.

[9] J. BERTIN, *Sémiologie graphique*, Paris, 1967 (réédition en 2005) : « La graphique tient ses lettres de noblesse de sa double fonction de mémoire artificielle et d'instrument de recherche. Outil rationnel et efficace lorsque les propriétés de la perception visuelle sont pleinement employées, elle fournit l'un des deux « langages » du traitement de l'information. L'écran cathodique lui ouvre un avenir illimité. », p. 6-8. Voir de même l'ouvrage classique de E.R. TUFTE, *The Visual Display of Quantitative Information*, Cheshire, 2001.

[10] C. LEVI-STRAUSS, « Les mathématiques de l'homme », in *Bulletin international des sciences sociales*, 6/4 (1954), p. 643-653, ici p. 647.

[11] La plupart des ouvrages se focalisent sur l'usage que font les historiens du visuel, et non sur les visuels sélectionnés ou produits par les historiens. Contrairement à ce que l'on observe en sciences dites dures, il semble qu'une telle archéologie n'ait jamais été tentée de façon systématique en sciences humaines. Voir néanmoins M. BANKS, *Visual Methods in Social Research*, Londres, 2001. En histoire médiévale, soulignons l'important essai de Bernhard Jussen, qui contient des éléments quantifiés sur les usages de l'image de Charlemagne par les historiens : B. JUSSEN, « Plädoyer für eine Ikonologie der Geschichtswissenschaft. Zur bildlichen Formierung historischen Denkens », in H. LOCHER (dir.), *Reinhart Koselleck. Politische Ikonologie. Perspektiven interdisziplinärer Bildforschung*, München, 2013, p. 260-279. On consultera aussi avec profit : L. GERVEREAU, D. MARECHAL, C. DELPORTE (éd.), *Quelle est la place des images en histoire ?*, Paris, 2008 ; J.-C. SCHMITT, « L'historien et les images », in O.G. OEXLE (éd.), *Der Blick auf die Bilder. Kunstgeschichte und Geschichte im Gespräch*, Göttingen, 1997, p. 7-51 (*Göttinger, Gespräche zur Geschichtswissenschaft*, 4) ; J.-P. TERRENOIRE, « Images et sciences sociales : l'objet et l'outil », in *Revue française de sociologie*, 26:3 (1985), p. 509-527. Concernant le visuel au Moyen Âge, nous renvoyons en dernier lieu à : J. BASCHET et P.-O. DITTMAR (éd.), *Les images dans l'Occident médiéval*, Turnhout, 2015.

Partant d'exemples isolés, puis d'un corpus bien circonscrit – les volumes annuels de la *Société des historiens médiévistes de l'enseignement supérieur (SHMESP)*[13] –, nous brosserons dans une première partie une typologie de l'utilisation des « images » en médiévistique[14]. Deux autres parties seront par la suite consacrées à des formes encore rares de représentations visuelles en histoire médiévale, celles à « vocation heuristique » : d'abord à partir de corpus d'objets (iconographiques, parchemins et édifices), puis de textes numérisés. L'hypothèse ici défendue est que l'extrême rareté des graphiques à vocation heuristique[15], et plus généralement des graphiques au sens large, en histoire (médiévale), est liée à la forme des rapports entretenus avec les documents anciens, ainsi qu'à la méthode historique au sens large, fondée pour partie sur la narration, le sujet et l'exemplarité. *In fine*, c'est l'intérêt de la « modélisation visuelle » en histoire qui sera discutée.

## 1. Établir une typologie : visuels, graphiques, modèles
### 1.1. Fonctions des « images » en médiévistique

À première vue, la variété des représentations graphiques en histoire médiévale est relativement importante. Elle va de l'iconographie à la carte, en passant par le schéma généalogique ou le *stemma* de manuscrits. Quels rôles remplissent les images dans le discours et l'entreprise analytiques et justificatrice des historiens[16] ? Les « graphiques » produits par les médiévistes constituent-ils des « données », des hypothèses, ou encore des modes d'exploration ? Peuvent-ils *faire preuve* ? Et en premier lieu, pour reprendre une interrogation de Catherine Allamel-Raffin[17], comment classer ces images scientifiques ? Plutôt que d'énumérer les différents types d'occurrences rencontrés dans les travaux dépouillés, au risque d'en oublier, il semble que l'on puisse distinguer ces « images » en termes de fonction et d'implication dans l'interprétation et la démonstration historienne, en lien avec le texte lui-même. Nous en distinguerons trois : (a) illustrer-embellir ; (b) situer-témoigner ; (c) explorer-modéliser.

---

[12] La notion de paradigme, parfois employée de façon abusive depuis quelques années, nous paraît ici particulièrement pertinente (nous détaillerons plus loin pourquoi). Concernant le concept, voir : T. KUHN, *La structure des révolutions scientifiques*, Paris, 1972 (publication originale en 1962) ; ID., *La tension essentielle. Tradition et changement dans les sciences*, Paris, 1990 (publication originale en 1977).

[13] Fondée en 1969 sous l'impulsion d'Edouard Perroy, la SHMESP réunit à la fois les historiens, archéologues et historiens de l'art travaillant en France sur le Moyen Âge. On trouvera des éléments pour établir son histoire dans : M. BALARD (éd.), *L'histoire médiévale en France. Bilan et perspectives*, Paris, 1991.

[14] Précisons d'emblée que le terme « image » est ici entendu au sens large, incluant toute mise en forme graphique ainsi que toute forme de reproduction iconographique.

[15] Autrement dit, des graphiques dont l'objectif est de permettre l'exploration de la documentation ancienne par la formalisation visuelle. Leur rôle est ainsi de « porter l'invisible au visible », pour reprendre l'expression de Monique Sicard (*La fabrique du regard [...]*, op.cit., p. 9). Sur cette question, nous renvoyons en premier lieu aux travaux de : F. MORETTI, *Graphes, cartes et arbres. Modèles abstraits pour une autre histoire de la littérature*, Paris, 2008 ; ID., *Distant Reading*, Londres-New York, 2013 ; E. NATALE, C. SIBILLE, N CHACHEREAU, P. KAMMERER, M. HIESTAND (dir.), *La visualisation des données en histoire / Visualisierung von Daten in der Geschichtswissenschaft*, Zürich, 2015.

[16] Nous mettons de côté la question politique, autrement dit le rôle de l'orientation idéologique des historiens dans leurs choix iconographiques. Sur ce thème, voir de nouveau à B. JUSSEN, « Plädoyer für eine Ikonologie der Geschichtswissenschaft. Zur bildlichen Formierung historischen Denkens », *op.cit.* En sciences expérimentales, voir Q. LADE, « Une belle image pour une bonne revue. Une ethnographie des représentations visuelles en sciences exéprimentales », in *Genèses*, vol. 103, 2016, p. 117-138.

(a). Illustrer : il s'agit là du premier degré de la présence iconographique. L'auteur présente un document dont le rapport avec le texte n'est pas nécessairement explicite, mais qui remplit un rôle d'embellissement, tout en étant, à divers degrés, lié au contenu de l'ouvrage ou au contexte historique évoqué dans celui-ci. C'est souvent le cas de la couverture du livre, dont les auteurs explicitent d'ailleurs rarement le choix, mais cette pratique est aussi fréquente dans les manuels. De nombreux ouvrages présentent en effet une enluminure, un manuscrit, un objet ou encore un bâtiment en frontispice, sans jamais revenir sur ce choix dans le corps du texte. Le visuel, par sa simple présence joue alors néanmoins un rôle : il place le lecteur en contact avec une documentation ancienne. Cette simple « présentation » aurait d'ailleurs certaines vertus, la présence du document ancien étant parfois réputée pour mettre en contact avec la société ancienne elle-même[18]. Ce rôle illustratif reste toutefois essentiellement esthétique : l'image est bien là pour embellir. Elle permet toutefois, dans certains cas, d'appuyer certaines pensées de l'auteur, tisser des liens entre différentes parties, sur un mode non-narratif.

(b). La seconde fonction du visuel en histoire médiévale est plus complexe et ses contours plus flous. Il s'agit de l'ensemble des éléments visuels sur lesquels s'appuie le discours historien. L'image fait alors, en quelque sorte, office de citation, de témoignage, de la même manière que l'historien cite un texte. Support de description, elle peut aussi appuyer une analyse, autrement dit contribuer à une démonstration – par exemple dans le cas du tableau de données –, mais encore donner un cadre : c'est l'exemple des cartes. Dans ce cas, le texte ne vient pas en appui de l'image, mais c'est à l'inverse le visuel qui renforce le discours historien. Proportionnellement, ces images sont majoritaires, et même de façon écrasante, puisqu'entrent selon nous dans cette catégorie[19] :

- Les documents iconographiques contenus dans les ouvrages d'histoire et d'histoire de l'art du Moyen Âge – dont la présence s'est évidemment accentuée grâce à la photographie[20] –, qu'il s'agisse d'édifices, de sculptures, de peintures (fresques ou

---

[17] C. ALLAMEL-RAFFIN, « La complexité des images scientifiques. Ce que la sémiotique de l'image nous apprend sur l'objectivité scientifique », *op.cit.*

[18] M. BLOCH, *Apologie pour l'histoire ou métier d'historien*, Paris, 1974, p. 51-52 (première édition en 1949) ; sur le rapport des historiens à leur documentation, nous renvoyons à J. MORSEL, « Les sources sont-elles 'le pain de l'historien' ? », in *Hypothèses 2003. Travaux de l'École doctorale d'histoire de l'Université Paris I Panthéon-Sorbonne*, 2004, p. 273-286 ; *ID.*, *La noblesse contra la ville ? Comment faire l'histoire des rapports entre nobles et citadins (en Franconie vers 1500) ?*, HDR inédit, Paris, 2009, p. 22-144.

[19] La question s'est posée de distinguer les images-objets (autrement dit les reproductions de documents anciens) des images produites par les historiens (graphes et graphiques, tableaux, schémas, etc.). Dans le cadre de cette analyse, cette distinction n'a pas semblé pertinente, car l'utilisation de ces deux « types » de visuel est plus ou moins identique. Dans les deux cas, ils sont en effet employés afin d'appuyer le discours historien, et donc à des fins démonstratives. Nous reviendrons plus loin sur cette question.

[20] Dont le développement en histoire, réel, demeure néanmoins limité : « On le sait, la photographie comme support d'informations et instrument de recherche n'est pas très valorisée dans nos sciences sociales, alors que d'autres disciplines comme la géologie, l'archéologie, l'astronomie ou même la médecine, reconnaissent dans l'image photographique un outil indispensable utilisé selon une méthodologie bien codifiée […] », dans A. PIETTE, « La photographie comme mode de connaissance anthropologique », in *Terrain. Anthropologie et sciences humaines*, 18 (1992), p. 129-136, ici p. 129. Au sein de sciences humaines et sociales, c'est probablement en anthropologie que la photographie joue le plus grand rôle : voir N. DIAS, « Photographier et mesurer : les portraits

tableaux), de manuscrits, de cartes, de gravures ou encore d'objets techniques, d'objets dits du quotidien et des dessins (par exemple en archéologie, histoire urbaine et histoire des techniques)[21]. Dans ces conditions, l'image n'est pas un outil, mais bien l'objet sur lequel porte l'analyse historienne[22]. Ces reproductions iconographiques, photographies ou dessins sont la plupart du temps placés dans un cahier central ou final, souvent pour des raisons éditoriales, ce qui induit un rapport particulier du texte à ces documents, qui sont souvent là pour justifier, témoigner. En histoire de l'art (et plus spécialement du Moyen Âge)[23], on peut faire remonter en France la *systématisation* de cette pratique à François Roger de Gaignières [1642-1715][24] et Jean-Baptiste Seroux d'Agincourt [1730-1814][25], puis à Henry Lemonnier [1842-1936][26] et Émile Mâle [1862-1954][27], selon différentes phases, en Allemagne à Anton Springer [1825-1891][28], Aby Warburg [1866-1929][29] et surtout de l'historien et *Provinzialkonservator* Paul Clemen [1866-1947][30]. Enfin, en Histoire médiévale, sur le versant états-unien, Arthur Kingsley Porter [1883-1933] a joué un rôle déterminant, en particulier à travers son ouvrage monumental, *Romanesque Sculpture of the Pilgrimage Roads*, paru en 1923[31].

---

anthropologiques », in *Romantisme*, 84 (1994), p. 37-49 ; C. LISSALDE, « L'image scientifique. Définitions, enjeux et questions », in *Bulletin des bibliothèques de France*, 46:5 (2001), p. 26-33, qui se concentre essentiellement sur la photographie ; S. PINK (éd.), *Doing Visual Ethnography*, Londres, 2001 ; M. SOUKUP, « Photography and drawing in Anthropology », in *Slovak Ethnology*, 62 (2014), p. 534-546.

[21] Là-encore, les anthropologues dominent le champ d'étude. Voir par exemple : H. GEISMAR, « Drawing It Out », in *Visual Anthropology Review*, 30:2 (2014), p. 97-113.

[22] J.-P. TERRENOIRE, « Images et sciences sociales : l'objet et l'outil », *op.cit.* ; J. BASCHET, « De l'usage des images en Histoire médiévale », in *Ménestrel*, collection *De l'usage de…*, Paris, 2012, http://www.menestrel.fr/spip.php?rubrique1691.

[23] Sur cette question, l'article de Wolfgang Freitag demeure une référence : W. FREITAG, « Early uses of photography in the history of art », in *Art Journal*, 39:2 (1979), p. 117-123 ; voir de même : C. WELGER-BARBOZA, « L'histoire de l'art et sa technologie. Concordance des temps », texte disponible en ligne sur le blog scientifique *L'Observatoire critique*, Paris, 2012 (https://observatoire-critique.hypotheses.org/1862) ; C. CARAFFA (éd.), *Photo Archives and the Photographic Memory of Art History*, Berlin-Munich, 2011.

[24] A. RITZ-GUILBERT, *La collection Gaignières. Un inventaire du royaume au XVIIe siècle*, Paris, 2016.

[25] J.-B. SEROUX D'AGINCOURT, *Histoire de l'art par les monuments, depuis sa décadence au IVe siècle jusqu'à son renouvellement au XVIe*, 6 vol., Paris, 1809-1823 ; nouvelle édition augmentée d'un volume, Turin, 2005. Concernant ses travaux, voir D. MONDINI, *Mittelalter im Bild : Seroux d'Agincourt und die Kunsthistoriographie um 1800*, Zurich, 2005.

[26] Concernant Henry Lemonnier, on consultera avec profit la notice très détaillée de L. THERRIEN, « Lemonnier, Henry (1842, Saint-Prix [Val d'Oise] – 1936, Paris) », in P. SENECHAL et C. BARBILLON (dir.), *Dictionnaire critique des historiens de l'art actifs en France de la Révolution à la Première Guerre mondiale*, Paris, 2009 (disponible sur le site de l'INHA : https://www.inha.fr).

[27] D. RUSSO, « Émile Mâle (1862-1954) : l'invention de l'iconographie historique », in *Comptes rendus des séances de l'Académie des Inscriptions et Belles-Lettres*, 148:4 (2004), p. 1641-1650 ; ID., « Cluny et la construction de l'histoire de l'art en France, d'Émile Mâle (1862-1954) à Henri Focillon (1881-1943) », in D. MEHU (dir.), *Cluny après Cluny. Constructions, reconstructions, et commémorations, 1790-2010*, Rennes, 2013, p. 197-208 ; ID., « Émile Male, l'art dans l'histoire », in *Émile Mâle, 1862-1954 : la construction de l'œuvre. Rome et l'Italie*, Rome, 2005, p. 251-271 ; à placer utilement en parallèle d'A. GRABAR, « Notice sur la vie et les travaux de M. Émile Mâle, membre de l'Académie », in *Comptes rendus des séances de l'Académie des Inscriptions et Belles-Lettres*, 106:2 (1962), p. 329-344 ; W. SAUERLANDER, « Mâle, Émile (2 juin 1862, Commentry – 6 octobre 1954, Fontaine-Chaalis) », in P. SENECHAL et C. BARBILLON (dir.), *Dictionnaire critique des historiens de l'art […]*, op.cit.

[28] J. RÖßLER, *Poetik der Kunstgeschichte. Anton Springer, Carl Justi und die ästhetische Konzeption der deutschen Kunstwissenschaft*, Berlin, 2009.

[29] A. WARBURG, *Der Bilderatlas Mnemosyne*, Berlin, 2000 ; ID., *L'Atlas mnémosyne*, Paris, 2012 ; G. DIDI-HUBERMAN, « Aby Warburg et l'archive des intensités », *Études photographiques*, 10 (2001), https://etudesphotographiques.revues.org/268 ;

[30] H. LÜTZELER, « Clemen, Paul », in *Neue Deutsche Biographie*, volume 3, Berlin, 1957, p. 281.

[31] A. KINGSLEY PORTER, *Romanesque Sculpture of the Pilgrimage Roads*, 10 volumes (dont 9 de planches), Boston, 1923 ; la photographie joue aussi un rôle central dans ID., *Medieval architecture: its origins and development, with lists of monuments and bibliographies*, New York, 1908 ; ID., *Lombard Architecture*, New

- Nous trouvons aussi (mais plus rarement) des cartes, dont l'objectif principal est souvent de déterminer le cadre géographique dans lequel se positionne le discours historien[32]. Elles se situent majoritairement en début ou en fin d'ouvrage, et ceci dès le XIXe siècle[33], tout comme le sont les analyses géographiques au sein des grandes monographies d'histoire régionale des années 1950-1980[34]. S'ajoute à cela les plans de sites archéologiques, qui sont toutefois un peu différent dans leur fonctionnement, puisqu'ils sont des documents créés par l'historien ou l'ingénieur lui-même.
- Des visuels de tous types : stemmas, schémas prosopographiques, plus rarement schémas de parenté, frises chronologiques, tableaux, courbes, diagrammes ou encore réseaux[35]. La plupart du temps, ces éléments arrivent en fin d'analyse, autrement dit en fin d'article, de chapitre ou d'ouvrage, voire en annexe. Ils ne sont donc pas, au sens strict, un support exploratoire, mais visent à synthétiser le discours historien et/ou à l'appuyer tout en le situant dans un cadre plus vaste.

(c). La troisième fonction potentielle des graphiques est de permettre l'exploration ou la modélisation des informations contenues dans les documents, afin de rendre possible leur analyse sous un jour nouveau. Une part plus ou moins importante des visuels placés dans la fonction précédente pourrait d'ailleurs parfaitement s'intégrer à celle-ci, puisqu'ils présentent un certain degré de formalisation : les stemmas, les réseaux, les schémas prospographiques, les cartes, etc. La plupart permettrait à l'historien, si ce dernier le souhaitait, de construire un discours fondé sur la modélisation graphique, à distance raisonnable du matériau documentaire[36], qu'il soit textuel, archéologique ou iconographique. Toutefois, ces exemples restent rares et, encore une fois, ces graphiques arrivent en fin d'ouvrage ou d'article, en synthèse, en non en début – ce qui ne permet pas de fonder le discours historien sur ces images, qui sont un point d'arrivée et non de départ pour l'analyse. Pour le dire autrement, tout se passe donc comme si la formalisation découlait du discours et non l'inverse.

Parallèlement, on peut aussi remarquer que ces trois fonctions – illustrer, situer, explorer – correspondent largement à trois degrés d'abstraction, et supposent des regards différentiés sur

---

Haven 1915-1917. Concernant Porter, voir : K. BRUSH, « Arthur Kingsley Porter and the Transatlantic Shaping of Art History, ca. 1910-1930 », in R. SUOMINEN-KOKKONEN (dir.), *The Shaping of Art History in Finland*, Helsinki, 2007, p. 129-142 ; *ID.*, « Arthur Kingsley Porter et la genèse de sa vision de Cluny », in D. MÉHU (dir.), *Cluny après Cluny [...]*, op.cit., p. 209-222.

[32] Nous distinguons ici les cartes-outils (carte de localisation ou carte d'analyse) des cartes-objets (plans et cartes anciennes), qui relèvent de la catégorie précédente.

[33] Pour un bilan de l'usage des cartes numériques en histoire, voir J.-L. ARNAUD, « Nouvelles méthodes, nouveaux usages de la cartographie et de l'analyse spatiale en histoire », in J.-P. GENET et A. ZORZI (dir.), *Les historiens et l'informatique : un métier à réinventer*, Rome, 2011, p. 199-220 ; A. GUERREAU, « Postface. Traitement des données historiques spatialisées. Que faire ? et comment ? », in M.-J. GASSE-GRANDJEAN et L. SALIGNY (dir.), *La géolocalisation des sources anciennes ?*, numéro hors-série du *Bulletin du Centre médiéval d'Auxerre* (n° 9), Auxerre, 2016, https://cem.revues.org/13840.

[34] Par exemple : G. DUBY, *La société aux XIe et XIIe siècles dans la région mâconnaise*, Paris, 2002 (première édition en 1953), p. 582-595, qui donne douze cartes ou plans, tous placés en fin d'ouvrage ; R. FOSSIER, *La terre et les hommes en Picard, jusqu'à la fin du XIIIe siècle*, Amiens, 1987, dans lequel on trouve une unique carte dépliante (mais d'assez nombreuses photographies), placée en fin de volume ; P. TOUBERT, *Les structures du Latium médiéval. Le Latium et la Sabine du IXe à la fin du XIIe siècle*, Rome, 1973, où les cartes dépliantes et détachables sont placées en fin de volume ; P. BONNASSIE, *La Catalogne au tournant de l'an mil*, Paris, 1990, se distingue néanmoins avec des cartes et plans situés à différents endroits du volume (par exemple, p. 26, 33, 44, 55, 104, 117, 138, 151, 159, 174, 226, etc.).

[35] Sur l'intérêt de ces visuels en Sciences humaines et sociales, nous renvoyons de nouveau à F. MORETTI, *Graphes, cartes et arbres. Modèles abstraits pour une autre histoire de la littérature*, op.cit.

[36] Cette distance « raisonnable » n'implique pas pour autant de se passer d'un retour aux documents, qui reste strictement nécessaire et doit toujours faire suite à une investigation formalisée. Nous développerons plus loin cet argument.

la documentation. En aucun cas toutefois, il ne faudrait définir une hiérarchie entre ces fonctions du visuel : elles sont complémentaires, à l'instar des méthodes qualitatives et quantitatives[37].

### 1.2 Fréquence et répartition du visuel dans un corpus scientifique (*SHMESP*)

Ces rapides remarques sur la nature des images employées par les médiévistes n'en disent toutefois pas long sur la pratique concrète des historiens. Afin de donner une assise plus ferme à ces observations empiriques, un corpus spécialisé a été dépouillé. Il s'agit des volumes collectifs de la *Société des Historiens médiévistes de l'Enseignement Supérieur Public*, qui paraissent annuellement depuis 1970[38]. Trente-sept de ces ouvrages collectifs sont numérisés sur le site *Persée*, couvrant les années 1970 à 2006, ce qui facilite grandement leur appréhension en tant que tout[39]. En définitive, cet ensemble représente un total de plus de 11 000 pages, pour près de 800 figures. Ces dernières ont été décomptées par année, en calculant un ratio du nombre de figures par rapport au nombre de pages[40], et en distinguant les différents types de graphiques.

Avant de présenter les résultats de cette enquête, il faut mentionner plusieurs biais liés au corpus : a) celui-ci comporte un plus grand nombre de visuels que la moyenne des ouvrages de médiévistique. En effet, puisqu'il s'agit de volumes collectifs, le nombre de cartes localisant les terrains d'enquête, ainsi que de visuels de synthèses a tendance à augmenter, influençant mécaniquement le ratio figures/pages ; b) la dimension collective des ouvrages introduit un biais typologique : les cartes et les tableaux de synthèse sont surreprésentées par rapport à d'autres types graphiques ; c) les thèmes retenus par les médiévistes influent fortement sur le nombre d'images présent. Ainsi, le volume de 1972 consacré à *La construction*[41] contient logiquement plus de plans que celui de 1999 sur *L'étranger*[42] ; d) les années 2007 à 2015 n'ont pas été couvertes par l'analyse. Cette situation regrettable s'explique pour partie par l'absence de numérisation des volumes sur Persée.

---

[37] Dès 1981, J.-P. Genet s'interrogeait sur la logique sous-jacente à cette opposition, J.-P. GENET, « Éditorial », in *Le médiéviste et l'ordinateur*, 5 (printemps 1981), p. 1 : « [L]'opposition quantitatif-qualitatif a-t-elle un sens ? ».
[38] La liste des volumes est la suivante : *La démographie médiévale. Sources et méthodes* (n° 1, 1970), in *Annales de la Faculté des Lettres et Sciences humaines de Nice*, 17 (1972) ; *Le vin au moyen âge : production et producteurs* (n° 2, 1971), Grenoble, 1978 ; *La construction au Moyen Âge. Histoire et archéologie* (n° 3, 1972), Paris, 1973 (*Annales littéraires de l'Université de Besançon*) ; *Les principautés au Moyen-Âge* (n° 4, 1973), Bordeaux, 1979 ; *Aspects de la vie conventuelle aux XIe-XIIe siècles* (n° 5, 1974), in *Cahiers d'histoire*, 20:2 (1985) ; *La mort au Moyen Âge* (n° 6, 1975), Strasbourg, 1977 ; *Les transports au Moyen Âge* (n° 7, 1976), in *Annales de Bretagne et des Pays de l'Ouest*, 85:2 (1978) ; *L'historiographie en Occident du Ve au XVe siècle* (n° 8, 1977), in *Annales de Bretagne et des Pays de l'Ouest*, 87:2 (1980) ; *Occident et Orient au Xe siècle* (n° 9, 1978), Paris, 1979 ; *Le paysage rural : réalités et représentations* (n° 10, 1979), in *Revue du Nord*, 62 (1980) ; *Le paysage urbain au Moyen-Âge* (n° 11, 1980), Lyon, 1981 ; *Les entrées dans la vie. Initiations et apprentissages* (n° 12, 1981), in *Annales de l'Est*, 5e série, 1-2, 1982 ; *Temps, mémoire, tradition au Moyen Âge* (n° 13, 1982), Aix-en-Provence-Marseille, 1983 ; *L'Église et le siècle de l'an mil au début du XIIe siècle* (n° 14, 1983), in *Cahiers de Civilisation Médiévale*, 27:1-2 (1984) ; *Le monde animal et ses représentations au Moyen Âge (XIe-XVe siècles)* (n° 15, 1984), Toulouse, 1985 ; *Les origines des libertés urbaines* (n° 16, 1985), Rouen, 1990 ; *L'Europe et l'Océan au Moyen Âge. Contribution à l'Histoire de la Navigation* (n° 17, 1986), Nantes, 1988 ; *Le combattant au Moyen Âge* (n° 18, 1987), Nantes, 1991 ; *Le marchand au Moyen Âge* (n° 19, 1988), Nantes, 1992 ; *L'histoire médiévale en France. Bilan et perspectives* (n° 20, 1989), Paris, 1991 ; *Villages et villageois au Moyen Âge* (n° 21, 1990), Paris, 1992 ; *Le clerc séculier au Moyen Âge* (n° 22, 1991), Paris, 1993 ; *Les princes et le pouvoir au Moyen Âge* (n° 23, 1992), Paris, 1993 ; *La circulation des nouvelles au Moyen Âge* (n° 24, 1993), Rome, 1994 ; *Miracles, prodiges et merveilles au Moyen Âge* (n° 25, 1994), Paris, 1995 ; *Voyages et voyageurs au Moyen Âge* (n° 26, 1995), Paris, 1996 ; *Les élites urbaines au Moyen Âge* (n° 27, 1996), Paris-Rome, 1997 ; *L'argent au Moyen Âge* (n° 28, 1997), Paris, 1998 ; *Les serviteurs de l'État au Moyen Âge* (n° 29, 1998), Paris, 1999 ; *L'étranger au Moyen Âge* (n° 30, 1999), Paris, 2000 ; *Le règlement de conflits au Moyen Âge* (n° 31, 2000), Paris, 2001 ; *Les échanges culturels au Moyen Âge* (n° 32, 2001), Paris, 2002 ; *L'expansion occidentale (XIe-XVe siècles), formes et conséquences* (n° 33, 2002), Paris, 2003 ; *Montagnes médiévales* (n° 34, 2003), Paris, 2004 ; *Ports maritimes et ports fluviaux au Moyen Âge* (n° 35, 2004), Paris, 2005 ; *Les villes capitales au Moyen Âge* (n° 36, 2005), Paris, 2006 ; *Construction de l'espace au Moyen Âge* (n° 37, 2006), Paris, 2007.
[39] http://www.persee.fr/
[40] Il s'agit tout simplement de diviser le nombre de figures au sein d'un volume, par le nombre de pages dans celui-ci.

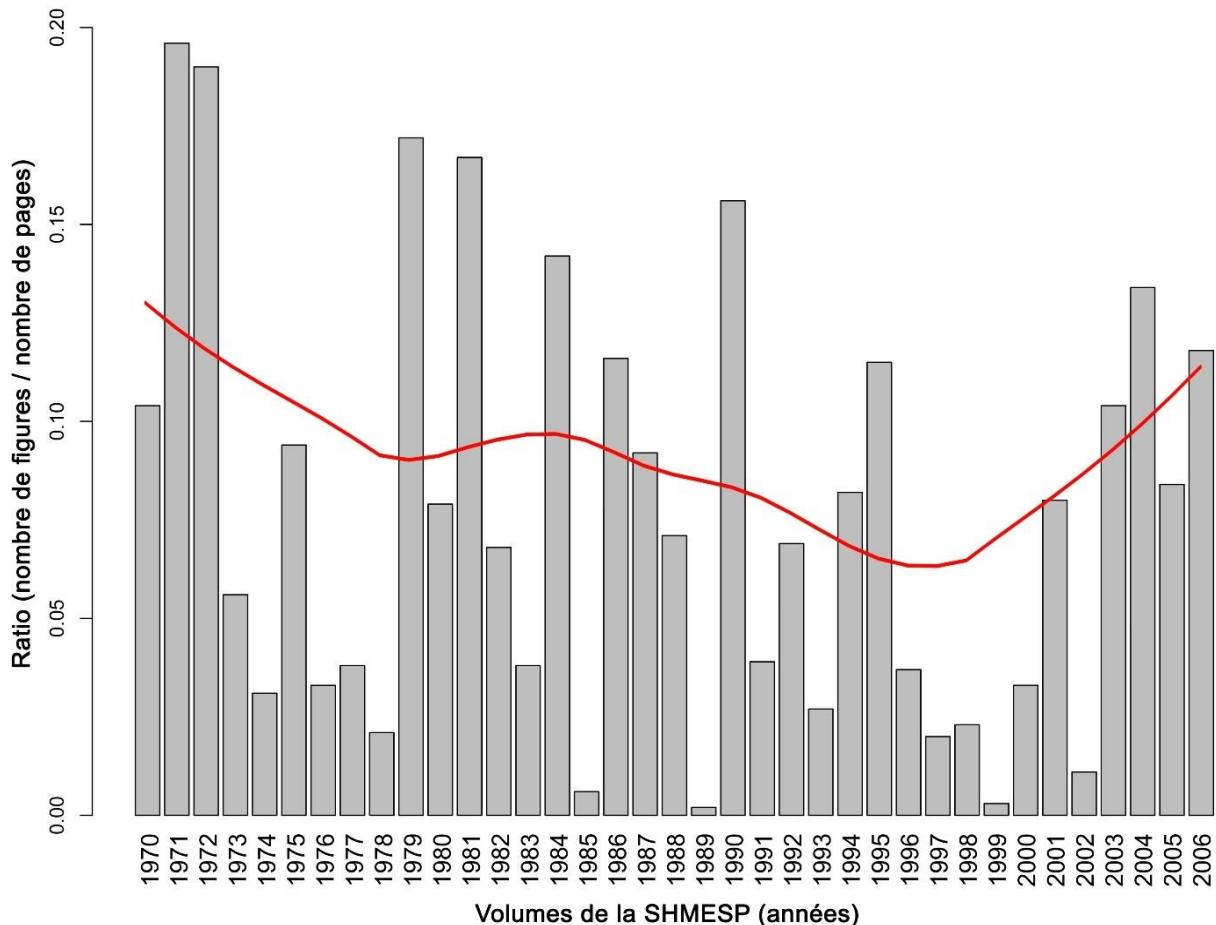

**Fig. 1 :** Évolution du nombre de figures dans les volumes de la SHMESP (1970-2006). La ligne rouge indique la tendance générale (lissage).

Ces points mentionnés, l'examen rapide du corpus mène à une première observation : 798 figures, tableaux compris, sur près de 11 000 pages au total, est un score relativement bas[43]. En calculant la moyenne puis la médiane des ratios figures/pages pour chaque année, nous obtenons respectivement un score de 0,08 et 0,07 : autrement dit, environ sept figures par tranche de cent pages. Si nous retirons les tableaux de ce décompte, nous obtenons une médiane de trois figures pour cent pages – tout en sachant que ce chiffre est plus élevé que la moyenne des productions de la discipline, pour les raisons précédemment évoquées.

Dans un second temps, une analyse globale du ratio figures/pages montre une tendance nette à la chute entre les années 1970 et 2000 (fig. 1). Non seulement les pics correspondant à des volumes renfermant de nombreuses images (comme le volume consacré au *Vin* en 1971, à

---
[41] *La construction au Moyen Âge. Histoire et archéologie*, op.cit.
[42] *L'étranger au Moyen Âge*, op.cit.
[43] Il s'agirait néanmoins de comparer ce score à celui que l'on pourrait obtenir sur des volumes en sociologie, psychologie ou anthropologie.

la *Construction* en 1972, au *Paysage rural* en 1979[44]) ont tendance à diminuer en importance, mais le nombre de volumes contenant très peu de représentations augmentent (*Les libertés urbaines* en 1985, *Le clergé séculier* en 1991, *L'argent* en 1997[45]). La tendance semble repartir à la hausse pour les années 2000-2006 et sans doute au-delà, mais les apparitions des dispositifs visuels restent rares.

Une lecture par type d'image permet d'en savoir plus sur cette tendance :

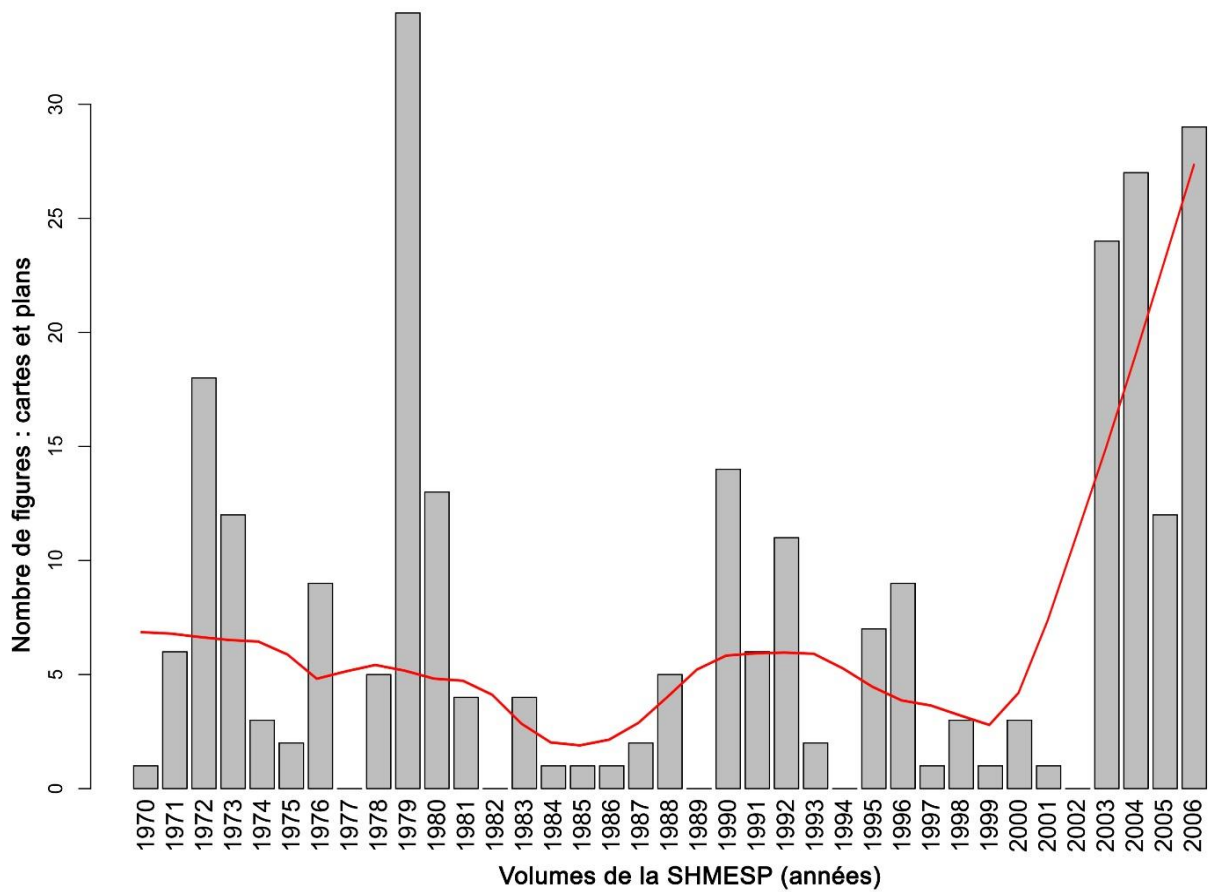

**Fig. 2 :** Évolution du nombre de cartes et de plans dans les volumes de la SHMESP (1970-2006).

Tout d'abord, au début du XXI[e] siècle, les cartes et les plans (271 occurrences au total, soit 34% du corpus, fig. 2) ont tendance à se généraliser. Certes, le volume de 1979 consacré au *Paysage rural*[46] contenait déjà de nombreux documents de ce type (34 au total), tout comme ceux consacré à la *Construction* et aux *Principautés*[47]. Cette tendance devient toutefois systématique dans les années 2000, sans doute sous la double influence des méthodes archéologiques (pour les plans) et des systèmes d'information géographique (pour les cartes au sens large)[48].

---

[44] *Le vin au moyen âge : production et producteurs*, op.cit. ; *La construction au Moyen Âge. Histoire et archéologie*, op.cit. ; *Le paysage rural : réalités et représentations*, op.cit.
[45] *Les origines des libertés urbaines*, op.cit. ; *Le clerc séculier au Moyen Âge*, op.cit. ; *L'argent au Moyen Âge*, op.cit.
[46] *Le paysage rural : réalités et représentations*, op.cit.
[47] *Les principautés au Moyen-Âge*, op.cit.

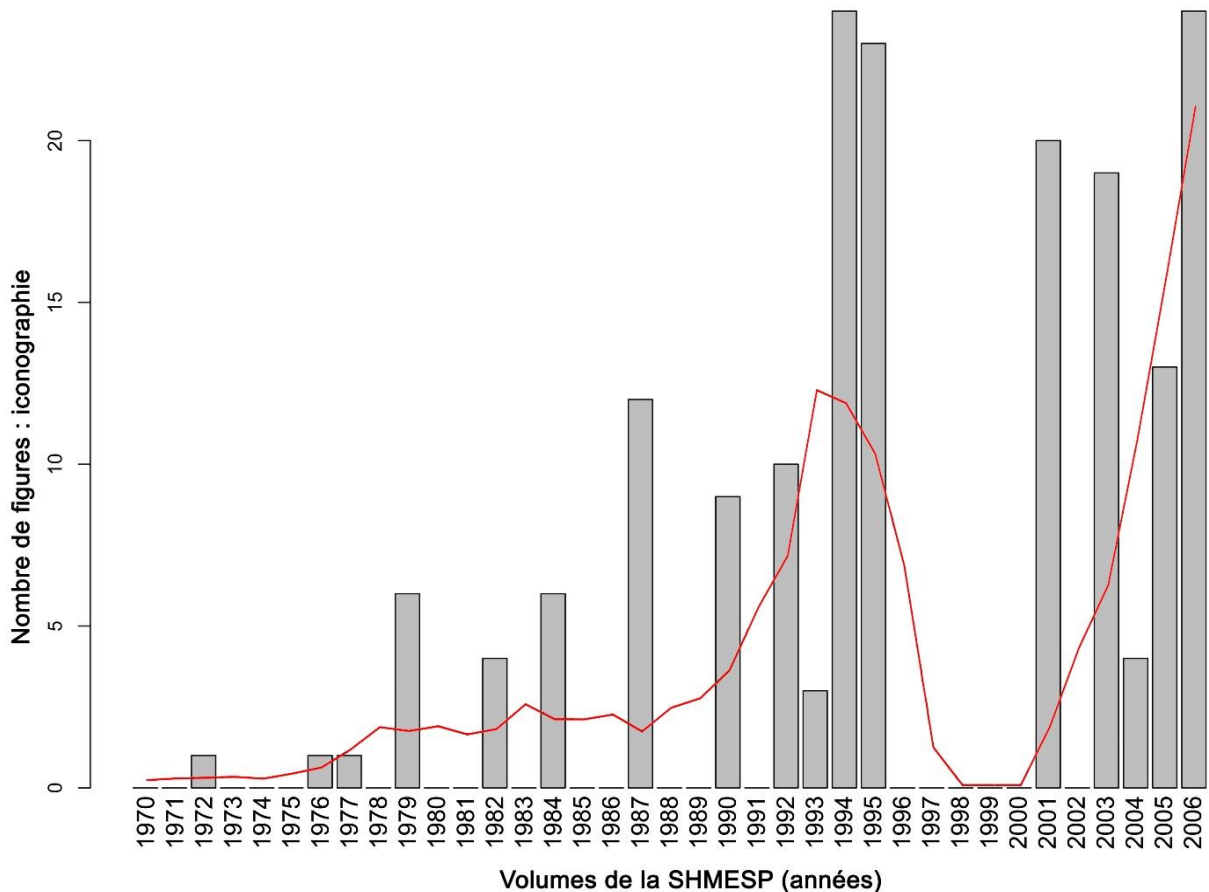

**Fig. 3 :** Évolution du nombre de documents iconographiques
dans les volumes de la SHMESP (1970-2006).

De façon similaire, on constate un intérêt de plus en plus net pour les documents iconographiques (180 occurrences, soit 22% du corpus, fig. 3). Cette seconde tendance se développe au milieu des années 1990, mais semble particulièrement appuyée dans les années 2000, indépendamment d'ailleurs du sujet retenu pour la conférence. Ce second phénomène est sans doute consécutif à la progressive prise au sérieux de l'impératif de Jacques Le Goff de diversifier les documents de l'historien[49], ainsi qu'à l'émergence d'une plus grande pluridisciplinarité.

---

[48] Voir les bilans proposés dans : J. CHAPELOT (dir.), *Trente ans d'archéologie médiévale en France. Un bilan pour un avenir*, Caen, 2010 ; I. CARTRON et L. BOURGEOIS, « Archéologie et histoire du Moyen Âge en France : du dialogue entre disciplines aux pratiques universitaires », in *Être historien du Moyen Âge au XXI$^e$ siècle*, Paris, 2007, p. 133-148.

[49] J. LE GOFF et P. TOUBERT, « Une histoire totale du Moyen Âge est-elle possible ? » in *Actes du 100$^e$ Congrès national des Sociétés savantes, section de philologie et d'histoire*, Paris, 1977, p. 31-44 ; P.TOUBERT, « Tout est document », in J. REVEL et J.-C. SCHMITT (dir.), *L'Ogre historien. Autour de Jacques Le Goff*, Paris, 1998, p. 85-105.

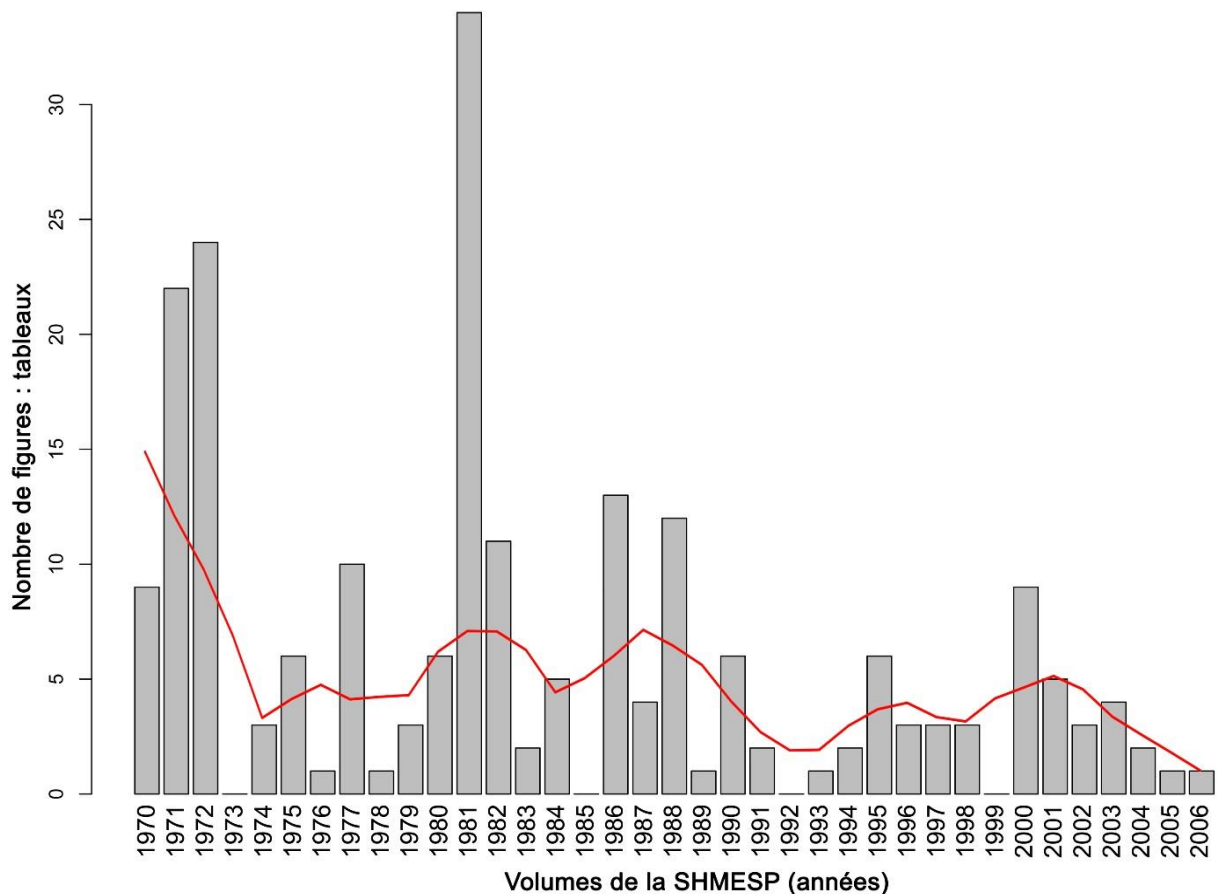
**Fig. 4 :** Évolution du nombre de tableaux dans les volumes de la SHMESP (1970-2006).

À l'inverse toutefois, les occurrences de tableaux (218 au total, soit 27% du corpus, fig. 4) ont tendance à baisser lentement depuis les années 1970, sans remonter au début du XXI[e] siècle. Cette troisième évolution peut probablement s'expliquer par une forme de malaise, croissant au cours de ces décennies, vis-à-vis des approches formalisées et quantitatives[50].

---

[50] A. GUERREAU, *L'avenir d'un passé incertain*, Paris, 2001, p. 115-131 (*Triomphe et évanouissement de l'« histoire quantitative »*) ; *ID.*, « L'étude de l'économie médiévale : genèse et problèmes actuels », in J. LE GOFF et G. LOBRICHON (dir.), *Le Moyen Âge aujourd'hui. Trois regards contemporains sur le Moyen Âge : histoire, théologie, cinéma*, Paris, 1998, p. 31-82.

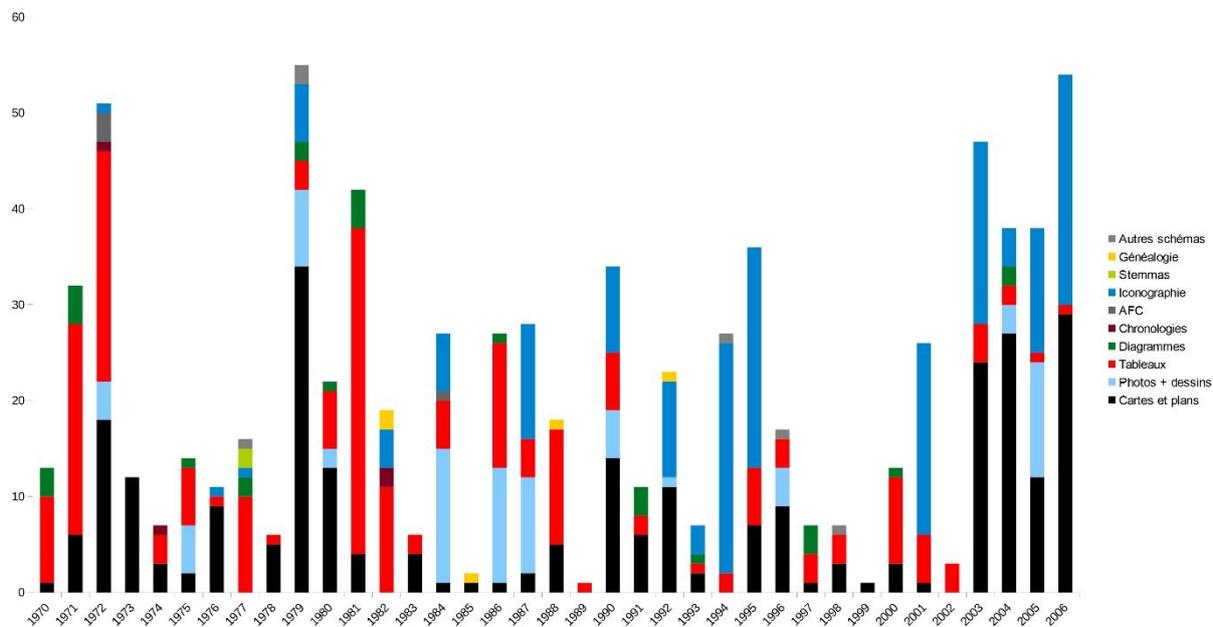
**Fig. 5 :** Évolution du nombre de figures dans les volumes de la SHMESP (1970-2006). Chiffres bruts (sans ratio par page) et répartition par type.

En regroupant les cartes, les plans, les tableaux et l'iconographie, on obtient un total de 669 occurrences, soit près de 83% du total des visuels contenus dans les volumes. Si l'on ajoute à cela les photographies (environ 10% du corpus), ces quatre types documentaires constituent à eux seuls près de 93% du total (fig. 5). Reste alors seulement 49 figures en 36 ans, soit 7% des occurrences, qui n'entrent pas dans ces catégories – soit à peine plus d'une figure par an présentant des chronologies, diagrammes, stemmas, généalogies, analyses multivariées et autres schémas. Or, si l'on revient aux catégories fonctionnelles distinguées précédemment, force est de constater que seuls ces graphiques pourraient, avec les cartes exploratoires, appartenir à la catégorie des supports permettant d'explorer ou de modéliser un phénomène historique. L'étude de ce corpus, qui devra toutefois être poursuivit à partir d'autres ensembles, permet ainsi d'établir l'hypothèse suivante : non seulement les visuels sont rares en histoire (médiévale), mais plus encore, leur typologie fonctionnelle les confine souvent à un rôle soit illustratif, autrement dit situationnel, au statut de témoignages[51].

### 1.3. L'absence du visuel : causes et conséquences supposées

Comment expliquer une telle désaffection pour les figures dans la discipline historique ? Avant d'en revenir à la quasi-absence des représentations visuelles à fonction heuristique, sur l'utilité et les possibilités pour dépasser cette « crise », il semblait important de revenir plus en détail sur les causes généralement supposées du blocage. Plusieurs points peuvent être évoqués :

- Il existerait en premier lieu un certain nombre de difficultés techniques. Les historiens maîtriseraient tout simplement assez mal les méthodes permettant de créer et de manipuler des images, encore moins des corpus visuels ou à des dispositifs à vocation heuristique. S'il est vrai que ce point ne saurait complétement être élué, en particulier à cause de la formation initiale des historiens[52], il semble que l'argument repose plus sur une réalité disciplinaire que sur de réelles contraintes techniques. Apprendre à maîtriser les fonctions de base d'un SIG, d'un langage de script, d'un logiciel de retouche photo, ou de traitements statistiques, ne prend en effet guère plus de quelques mois[53].

---

[51] En ce sens, leur rôle est proche de celui des citations. On trouvera différentes réflexions sur ce thème dans : A. COMPAGNON, *La seconde main ou le travail de la citation*, Paris, 1978 ; A. GRAFTON, *Les origines tragiques de l'érudition. Une histoire de la note en bas de page*, Paris, 1998 ; P. BOUCHERON, « De l'usage des notes de bas de page en Histoire médiévale », in *Ménestrel. Médiévistes sur le net : sources, travaux et références en ligne*, collection *De l'usage de…*, Paris, 2012, en ligne : http://www.menestrel.fr/spip.php?rubrique1309.

- La nouveauté de ces méthodes ne peut pas non plus être un critère discriminant : certaines bases de données et autres traitements quantitatifs permettant la fabrication d'images exploratoires sont parfois disponibles depuis des décennies[54]. Il en va ainsi des analyses factorielles, découvertes au milieu des années 1960 par Jean-Paul Benzécri [1932-][55], dont le potentiel a été employé en psychologie ou en sociologie[56], mais plus rarement en histoire[57]. Pour la période avant 2007, les volumes de la SHMESP ne renferment par exemple que quatre graphes de ce type, dont trois au sein d'un même article[58]. Quant aux bases de données, qu'elles soient iconographiques ou textuelles,

---

[52] Le paysage disciplinaire évolue néanmoins lentement, avec l'introduction de formations aux méthodes visuelles, statistiques et informatiques. Le cas de l'École nationale des chartes est ici exemplaire, avec la création du Master « Technologies numériques appliquées à l'histoire », dont le succès est considérable.

[53] Les travaux doctoraux menés sous la direction ou avec l'aide de Daniel Russo, Eliana Magnani, Alain Guerreau ou encore Jean-Philippe Genet en sont la preuve. S'il nous est permis d'évoquer ici notre expérience personnelle, nous avons appris, au cours de notre thèse, à maîtriser les bases d'un SIG, plusieurs langages de script et différents outils d'analyse statistique. Certes, nous disposions d'un bagage solide en matière de numérique – mais c'est aujourd'hui le cas d'un nombre croissant d'étudiants, en particulier ceux en provenance des filières scientifiques.

[54] Ce point avait déjà été souligné dans : N. PERREAUX, « De l'accumulation à l'exploitation ? Expériences et propositions pour l'indexation et l'utilisation des bases de données diplomatiques », in A. AMBROSIO, S. BARRET et G. VOGELER (dir.), *Digital diplomatics. The computer as a tool for the diplomatist?*, Köln-Weimar-Wien, 2014, p. 187-210 (*Archiv für Diplomatik. Schriftgeschichte Siegel- und Wappenkunde*, Beiheft 14).

[55] J.-P. BENZECRI et al., *L'Analyse des données, 1 : La taxinomie ; 2 : L'Analyse des correspondances*, Paris, 1973 ; *ID.*, (dir.), *Pratique de l'analyse des données*, 4 volumes, Paris, 1980-1986.

[56] L'exemple le plus fameux étant bien entendu celui de P. BOURDIEU, *La distinction, critique sociale du jugement*, Paris, 1979 ; voir aussi : P. BOURDIEU et J.-C. PASSERON, *Les Héritiers. Les étudiants et la culture*, Paris, 1964.

[57] Avec quelques exceptions notables, en particulier dans les volumes de la revue *Le médiéviste et l'ordinateur*, au cours des années 1970-1980. Aujourd'hui, quelques médiévistes reprennent l'usage de ces méthodes avec succès : A. MAIREY, *Une Angleterre entre rêve et réalité. Littérature et société dans l'Angleterre du XIVe siècle*, Paris, 2007 ; S. LEPAPE, *Représenter la parenté du Christ et de la Vierge : l'iconographie de l'Arbre de Jessé en France du Nord et en Angleterre, du XIIIe siècle au XVIe siècle*, Paris, 2007 ; I. GUERREAU, *L'auto-représentation du clergé saxon au Moyen-Âge d'après les sceaux*, Paris, 2009 (désormais édité sous le titre *Klerikersiegel der Diözesen Halberstadt, Hildesheim, Paderborn und Verden im Mittelalter*, Hannover, 2013) ; R. MARCOUX, *L'espace, le monument et l'image du mort au Moyen Âge. Une enquête anthropologique sur les tombeaux médiévaux de la Collection Gaignières*, Québec, 2013 (thèse inédite) ; N. PERREAUX, « L'écriture du monde (I). Les chartes et les édifices comme vecteurs de la dynamique sociale dans l'Europe médiévale (VIIe-milieu du XIVe siècle) », in *Bulletin du Centre médiéval d'Auxerre*, 19.2 (2015), http://cem.revues.org/14264 ; *ID.*, « L'écriture du monde (II). L'écriture comme facteur de régionalisation et de spiritualisation du *mundus* : études lexicales et sémantiques », in *Bulletin du Centre médiéval d'Auxerre*, 20.1 (2016), http://cem.revues.org/14452 ; *ID.*, « Mesurer un système de représentation ? Approche statistique du champ lexical de l'eau dans la Patrologie latine », in *Mesure et histoire médiévale (XLIIIe congrès de la SHMESP, Tours, 31 mai-2 juin 2012)*, Paris, 2013, p. 365-374 ; J.-B. CAMPS, « Troubadours et analyses factorielles : approches statistiques à la représentation de l'auteur dans les chansonniers occitans A, I et K », in *Nouvelle Recherche en domaine occitan : approches interdisciplinaires*, Albi, 2009, http://halshs.archivesouvertes.fr/halshs-00824016.

[58] On trouvera trois cartes factorielles dans : M. BUR, « Essai de typologie de l'habitat seigneurial dans l'Argonne aux XIe-XIIe siècles d'après des vestiges relevés sur le terrain », in *La construction au Moyen Âge. Histoire et archéologie*, op.cit., p. 145-174 (ici p. 172-174). Le second article, qui contient ce que nous supposons être une analyse factorielle, ne précise pas la nature réelle du graphique : M. PASTOUREAU, « Quel est le roi des animaux ? », in *Le monde animal et ses représentations au Moyen Âge (XIe-XVe siècles)*, op.cit., p. 133-142 (ici p. 142). Que ces cartes analytiques apparaissent entre 1970 et 1985 correspond bien à l'essor de ces méthodes lors de cette période, et à leur recul lors des décennies suivantes. Hors de notre corpus, le volume *Mesure et histoire médiévale*, Paris, 2013 (congrès de Tours, 31 mai-2 juin 2012), auquel nous avons eu la joie de participer, contient trois cartes factorielles : J.-P. GENET, « Les théologiens parisiens : une approche par l'analyse factorielle », p. 135-

certaines existent depuis des décennies : ainsi, le *Corpus Thomisticum*, dont l'*Index* (*Thomisticus*) réalisé en collaboration par le père Busa et la société IBM a débuté en 1949[59].
- Un troisième argument, peut-être plus sérieux, concerne les modes de diffusion du savoir historien. L'impression de graphiques, en particulier de documents iconographiques, de cartes ou d'analyses statistiques complexes, demande un savoir-faire et implique un coût de fabrication pour des volumes papiers. Les impressions couleurs sont souvent de rigueur, engendrant des frais conséquents. Parallèlement, il faut admettre que la création de nombreuses revues en ligne favorise aujourd'hui la présence d'images scientifiques chez les médiévistes. Ainsi, l'exemple du *Bulletin du Centre médiéval d'Auxerre* est assez éloquent[60]. Malgré cette augmentation de la place du visuel dans les productions historiens, les représentations à vocation heuristique restent cependant très minoritaires, y compris dans ces revues numériques, à l'inverse des documents iconographiques, qui tendent à se multiplier.

## 2. Du visuel au modèle : passage(s)
### 2.1. L'historien, le corpus, la ruine

La désaffection pour les images exploratoires ou modélisatrice ne peut donc s'expliquer par des contingences strictement techniques. Leur quasi-absence est constante, sur le temps long, en dépit des progrès considérables du système-technique. L'hypothèse proposée ici sera est donc d'une autre nature. Les visualisations à vocation heuristique diffèrent en effet la méthode historienne classique sur trois plans[61] :

En premier lieu, la discipline est traditionnellement fondée, comme de nombreuses sciences humaines et sociales, sur un paradigme logocentrique. La linéarité du discours, la narration comme modèle explicatif, implique ainsi une mise en forme particulière de l'analyse, dans laquelle le recours aux procédés visuels reste le plus souvent dans l'ordre situationnel, du témoignage. On pointe ainsi vers des occurrences, qui renforcent le discours historien par leur simple présence et leur exemplarité. A l'inverse, le fait de placer un dispositif visuel au départ de la chaine d'analyse déconstruit la linéarité et le modèle explicatif de l'historien, du moins le réoriente. Lorsque sa vocation est heuristique, le graphique fait en effet apparaître des structures, c'est-à-dire des éléments centraux pour tout historien (y compris dans la narration), mais qui sont extrêmement délicats à décrire par le langage.

Deuxièmement, le rapport que l'historien entretient avec le document, et en particulier le document original, est (ou serait) lui aussi transformé par l'emploi d'outils heuristiques. En prétendant que certaines structures historiques ne peuvent repérées que par l'addition d'un filtre visuel-quantitatif, permettant de déconstruire puis de reconstruire de façon synthétisée les relations présentes dans documents anciens, les méthodes heuristiques défient le paradigme consistant à considérer que plus la proximité « matérielle » avec les documents est grande, plus l'analyse historienne est scientifiquement fondée[62]. Autrement dit : s'il y a besoin d'un filtre, c'est que le document ne parle pas de lui-même, ce qui représente un changement de perspective radical.

---

152 (ici p. 147 et 149) et N. PERREAUX, « Mesurer un système de représentation ? Approche statistique du champ lexical de l'eau dans la Patrologie latine », p. 365-374 (ici p. 370).

[59] Aujourd'hui http://www.corpusthomisticum.org/.

[60] Voir les volumes en ligne : https://cem.revues.org/.

[61] Il ne s'agit toutefois en aucun cas de prétendre que ces méthodes numériques « remplacent » les méthodes classiques. Tout au contraire, elles viennent s'y ajouter et permettent même d'y retourner, après la production de résultats nouveaux. En aucun cas, il ne peut donc y avoir substitution et ceci d'autant plus que la consultation des documents demeure toujours strictement nécessaire en cas de résultats « à distance » apparemment intéressant. Sur ces questions, voir F. MORETTI, *Graphes, cartes et arbres. Modèles abstraits pour une autre histoire de la littérature*, op.cit. ; *ID.*, *Distant Reading*, op.cit.

Troisièmement, la manière dont l'historien considère son corpus documentaire influence fortement la possibilité, ou non, de réaliser des visualisations à objectif heuristique. Les documents de l'antiquisant ou du médiéviste sont en effet réputées pour avoir subies de profondes transformations, entre leur production et leur conservation actuelle – ce qui est évidemment le cas à des échelles fines. Mais le paradigme historien va souvent plus loin en affirmant que les objets hérités du passé ne forment pas toujours des structures représentatives, en tant que *corpus*[63]. Cette hypothèse possède un corolaire pour les approches visuelles : si les « traces » des sociétés anciennes ne forment pas des ensembles cohérents, il n'est pas nécessaire d'étudier quantitativement leur formation et leur dynamique en tant que tout, puisque celles-ci seront de toute façon biaisées par les destructions successives, arbitraires ou non – en tout cas insaisissables. C'est ce que nous avons proposé d'appeler le « paradigme de la ruine »[64]. A l'inverse, l'emploi de visualisations heuristiques implique d'attribuer de façon a priori du sens et de la cohérence aux documents ainsi qu'aux corpus que ceux-ci forment[65].

Ayant évoqué la typologie et la fréquence des graphes en histoire médiévale, ainsi que certaines causes de leur absence, il s'agit maintenant de présenter quelques exemples montrant leur intérêt potentiel. La mise à distance des documents médiévaux par la visualisation est aujourd'hui en effet facilitée par une triple conjoncture, de plus en plus affirmée : 1. L'existence de très nombreuses bases de données et d'outils désormais adaptés pour les manipuler[66]. 2. D'autre part, le fait qu'une part croissante des historiens, en partie sous l'influence de l'anthropologie, ont conscience qu'il n'est ni possible ni souhaitable d'essayer d'interpréter immédiatement les documents anciens, qui sont frappés par l'altérité. Ils rejoignent ainsi, les précurseurs que furent Jacques Le Goff, Jean Pierre-Vernant et Marcel Détienne pour l'histoire ancienne et médiévale[67]. 3. Enfin, des expériences dans des domaines connexes, par exemple les travaux désormais célèbres de Franco Moretti[68], ont montré ce qu'une lecture distante des textes pouvait apporter, mais aussi la linguistique de corpus[69].

---

[62] Autrement dit la connaissance « directe » de la « trace » est considérée comme plus riche qu'une connaissance « indirecte ». Nous ne contestons pas que la consultation des manuscrits, lorsqu'ils subsistent, est d'une importance capitale. Simplement que ce mode de consultation de la documentation historique supplante (voire annule) tous les autres. Sur ces considérations, voir J. MORSEL, « Traces ? Quelles traces ? Réflexions pour une histoire non passéiste », in *Revue historique*, n° 680 (2016), p. 813-868 ; ID., « Du texte aux archives : le problème de la source », in E. Magnani (dir.), *Le Moyen Âge vu d'ailleurs*, Dijon, 2008 (Bulletin du Centre médiéval d'Auxerre, Hors-série n° 2), https://cem.revues.org/4132 ; ainsi que É. ANHEIM, « Singulières archives. Le statut des archives dans l'épistémologie historique. Une discussion de *La Mémoire, l'histoire, l'oubli* de Paul Ricoeur », dans *Revue de Synthèse*, n° 125 (2004), p. 153-182.
[63] Le cas des documents diplomatiques et des édifices des X$^e$-XII$^e$ siècles sont probablement paradigmatiques. Voir N. PERREAUX, « L'écriture du monde (I). Les chartes et les édifices comme vecteurs de la dynamique sociale dans l'Europe médiévale (VII$^e$-milieu du XIV$^e$ siècle) », *op.cit.* On peut néanmoins faire l'hypothèse que cette structuration (relative) des corpus anciens conservés serait valable pour de nombreux autres ensembles.
[64] L'usage de l'expression est présenté dans : N. PERREAUX, « Des structures inconciliables ? Cartographie comparée des chartes et des édifices « romans » (X$^e$-XIII$^e$ siècles) », in M.-J. GASSE-GRANDJEAN et L. SALIGNY (dir.), *Géolocalisation et sources anciennes ?*, Dijon, 2016 (Bulletin du Centre médiéval d'Auxerre, Hors-série n° 9), https://cem.revues.org/13817.
[65] La journée d'étude « Qu'est-ce qu'un corpus ? » (7 novembre 2016, Paris, IRHT), organisé par E. Magnani, fut l'occasion de discuter de ces problématiques.
[66] Précieux rappels, dans A. GUERREAU, « Pour un corpus de textes latins en ligne », in *Bulletin du Centre médiéval d'Auxerre* (Collection CBMA), Dijon, 2011, https://cem.revues.org/11787.
[67] J. LE GOFF, *La civilisation de l'Occident médiéval*, Paris, 1964 ; J.-P. VERNANT, *Mythe et pensée chez les Grecs. Études de psychologie historique*, Paris, 1965 ; ID., *Mythe et société en Grèce ancienne*, Paris, 1974 ; M. DETIENNE, *Crise agraire et attitude religieuse chez Hésiode*, Paris, 1963 (*Latomus*, 68).

Afin de présenter ces perspectives, et toujours dans l'idée d'exemplifier le rôle que pourraient avoir les visuels heuristiques en histoire (médiévale), les cas qui suivent seront principalement issus de nos recherches. Ce choix est avant tout pratique : d'autres chercheurs auraient pu contribuer à cette section, sans doute plus utilement que nous, en particulier Jean-Philippe Genet, Aude Mairey, Isabelle Guerreau, Séverine Lepape ou encore Robert Marcoux.

### 2.2. Visualiser la production diplomatique européenne ?

La première méthode de visualisation présentée combine analyses multivariées et projections cartographiques. Dans le cadre d'une thèse soutenue en 2014[70], nous avons constitué un corpus documentaire regroupant l'ensemble des documents diplomatiques du Moyen Âge numérisés à ce jour, soit 140 000 chartes. Nous souhaitions alors employer ce corpus afin de réaliser des enquêtes lexicographiques et sémantiques. Toutefois, le problème de sa représentativité a rapidement émergé. Si des tendances scripturaires étaient découvertes à partir de celui-ci, dans quelle mesure étaient-elles le reflet de dynamiques socio-historiques et non du corpus lui-même ? Afin d'affronter cette question, nous avons choisi de placer le corpus des textes diplomatiques au cœur de l'enquête, plutôt qu'un problème ou une thématique retenue arbitrairement[71].

Les chartes étaient en effet réputées pour avoir connu des destructions sélectives, pour des raisons mémorielles, « stratégique » et « idéologique » si l'on préfère, de la part des moines eux-mêmes[72]. Issue d'une réflexion émergeant lors des années 1990, cette perspective extrêmement fructueuse pour des échelles d'analyse fines, permettant de saisir certains enjeux historiques locaux, faisait-elle toujours sens à l'échelle d'un corpus européen ?

---

[68] F. MORETTI, *Graphes, cartes et arbres. Modèles abstraits pour une autre histoire de la littérature*, op.cit. ; *ID.*, *Distant Reading*, op.cit.

[69] Voir en premier lieu A. LÜDELING et M. KYTÖ (éd.), *Corpus Linguistics: An International Handbook*, 2 volumes, Berlin-New York, 2008.

[70] *L'écriture du monde. Dynamique, perception, catégorisation du mundus au moyen âge (VII$^e$-XIII$^e$ siècles). Recherches à partir de bases de données numérisées*, Dijon, 2014, réalisée sous la direction conjointe de Daniel Russo et d'Eliana Magnani. La thèse est actuellement en cours d'édition, dans la *Collection d'études médiévales de Nice* (Turnhout, Brepols).

[71] Ce qui renvoie bien entendu à l'idée d'« histoire-problème ». Attachée à l'École des Annales et en particulier à Lucien Febvre, la notion constituait toutefois, en son temps, un progrès important. En particulier parce qu'en plaçant le regard des historiens par-delà les faits, elle permettait de rejeter l'hypothèse d'une transparence des documents anciens, déjà évoquée. Voir J.-C. SCHMITT, « ''Façon de sentir et de penser''. Un tableau de la civilisation ou une histoire-problème ? », in H. ATSMA et A. BURGUIERE (dir.), *Marc Bloch aujourd'hui. Histoire comparée et sciences sociales*, Paris, 1990, p. 407-418. Cette tension entre « histoire-problème » et approche « corputielle »-quantitative avait déjà été soulignée dans B. LEPETIT, « L'histoire quantitative : deux ou trois choses que je sais d'elle », in *Histoire & Mesure*, vol. 4 (1989), p. 191-199, ici p. 195.

[72] Cette hypothèse nous paraît néanmoins valable à des échelles fines, et tous les médiévistes savent que les remaniements documentaires-archivistiques sont légions. Sur cette question, voir en premier lieu P. GEARY, « Entre gestion et *gesta* », in O. GUYOTJEANNIN, L. MORELLE et M. PARISSE (dir.), *Les cartulaires, Actes de la Table ronde organisée par l'École nationale des chartes et le GDR 121 du CNRS, Paris, 5-7 décembre 1991*, Paris, 1993, p. 13-27 ; *ID.*, *Mémoire et oubli à la fin du premier millénaire*, Paris, 1996 ; P. CHASTANG, *Lire, écrire, transcrire. Le travail des rédacteurs de cartulaires en Bas-Languedoc (XI$^e$-XIII$^e$ siècles)*, Paris, 2001 ; *ID.*, « Cartulaires, cartularisation et scripturalité médiévale : la structuration d'un nouveau champ de recherche », in *Cahiers de Civilisation Médiévale*, numéro consacré à *La médiévistique au XX$^e$ siècle. Bilan et perspectives*, n°49, Janvier-Mars 2006. p. 21-31 ; J. Morsel, *La noblesse contre la ville ? Comment faire l'histoire des rapports entre nobles et citadins (en Franconie vers 1500) ?*, Paris, 2009 ; W.C. BROWN, M.J. COSTAMBEYS, M.J. INNES et A.J. KOSTO (dir.), *Documentary culture and the laity in the early Middle Ages*, Cambridge, 2013 ; P. BERTRAND, *Les écritures ordinaires. Sociologie d'un temps de révolution documentaire (entre royaume de France et Empire, 1250-1350)*, Paris, 2015.

Deux critères simples ont été retenus dans un premier temps : la chronologie des actes et leur provenance. Nous avons ainsi décompté les documents édités, parfois simplement conservés, afin d'en réaliser une analyse distributionnelle (fig. 6). Il a ainsi été possible de montrer que les profondes disparités entre Bourgogne septentrionale ou Bourgogne méridionale ne pouvaient se réduire à des effets de conservation ou des influences aléatoires[73]. Dans cette région, il existe une opposition radicale entre les espaces méridionaux, incluant un très grand nombre d'actes pour les X$^e$-XI$^e$ siècles (entre autres ceux de Cluny, mais pas uniquement), et d'autre part une Bourgogne du Nord fragmentée en deux profils, centrés sur les XII$^e$ et XIII$^e$, voire XIV$^e$ siècle (fig. 7)[74].

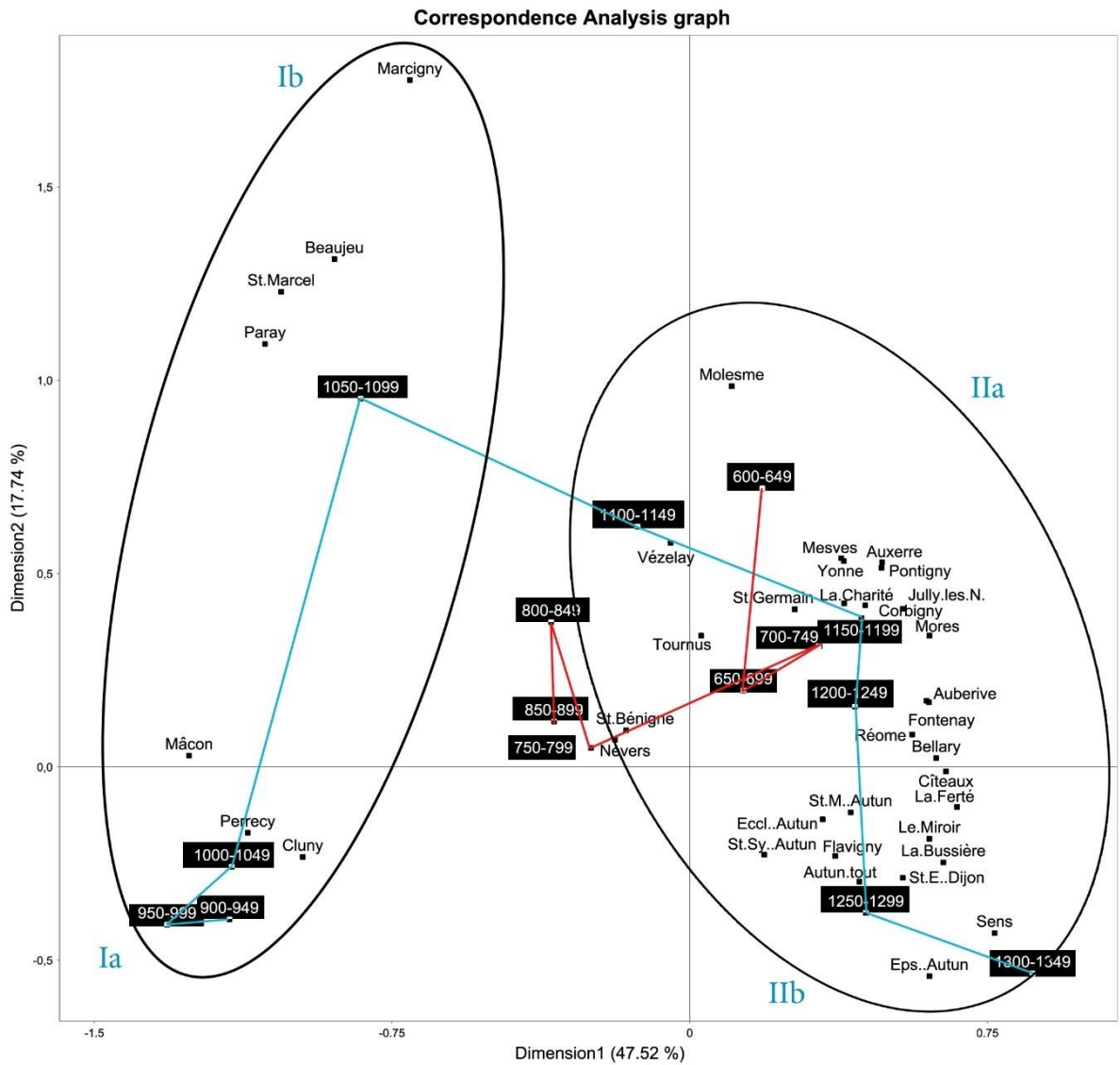

**Fig. 6 :** Corpus diplomatiques bourguignons, analyse factorielle du nombre d'actes produits par établissement, par demi-siècle (VII$^e$-milieu du XIV$^e$ siècle). Plan factoriel 1-2.

---

[73] Cette opposition entre Bourgogne du nord et Bourgogne du sud, avait déjà été constatée par André Déléage à d'autres niveaux : A. DELEAGE, *La Vie rurale en Bourgogne jusqu'au début du onzième siècle*, 3 volumes, Mâcon, 1941.

[74] L'application d'un algorithme de partitionnement (clustering) aux analyses factorielles permet de dégager trois groupes, reportés ici sur une première carte.

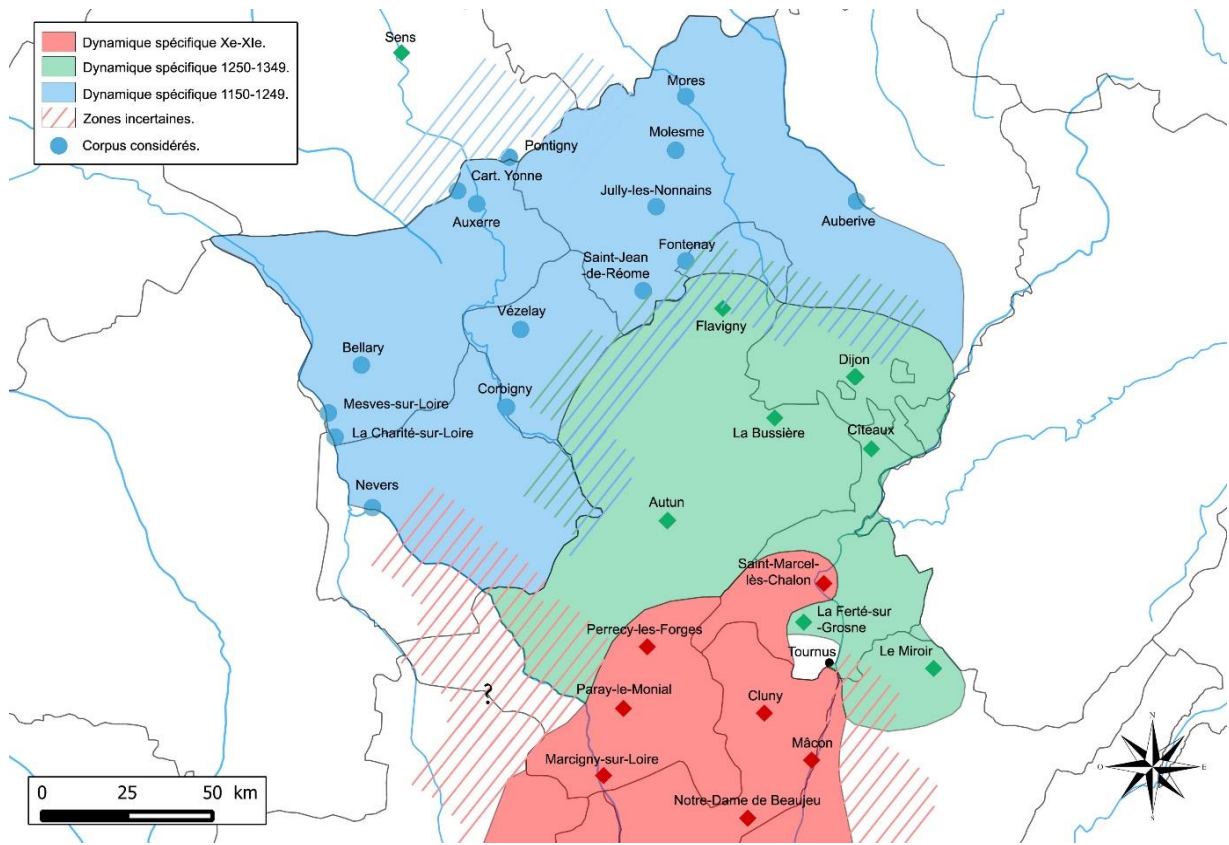

**Fig. 7 :** Corpus diplomatiques bourguignons, projection cartographique des analyses multivariées (cf. figure précédente).

Les investigations ont été poursuivies par la suite à une échelle plus vaste, selon un modèle que l'on a choisi de nommer « chrono-géographique ». Les dépouillements se sont donc étendus à la totalité des éditions d'actes diplomatiques disponibles pour l'actuelle France, d'une part pour les originaux, d'autre part pour les copies (fig. 8)[75]. Il apparaît alors que les structures repérées pour la Bourgogne sont aussi visibles à cette échelle, et que l'actuelle France est globalement scindée en deux dynamiques scripturaires, du moins en diplomatique, opposant un sud précoce (880-1150) et un nord plus tardif (1150-1300). La seconde constatation dégagée est qu'il n'existe pas d'opposition franche entre les originaux et les copies, en termes de distribution chrono-géographique (fig. 9)[76]. Deux hypothèses qu'il n'aurait pas été possible de proposer sans les méthodes de formalisation, de visualisation et en définitive de modélisation auxquelles nous avons eu recours.

---

[75] Soit respectivement environ 5 000 actes, puis 580 éditions et 147 000 actes.
[76] Présentation détaillée de ces hypothèses dans N. PERREAUX, « L'écriture du monde (I). Les chartes et les édifices comme vecteurs de la dynamique sociale dans l'Europe médiévale (VII[e]-milieu du XIV[e] siècle) », *op.cit.* ; *ID.*, « L'écriture du monde (II). L'écriture comme facteur de régionalisation et de spiritualisation du *mundus* : études lexicales et sémantiques », *op.cit.*

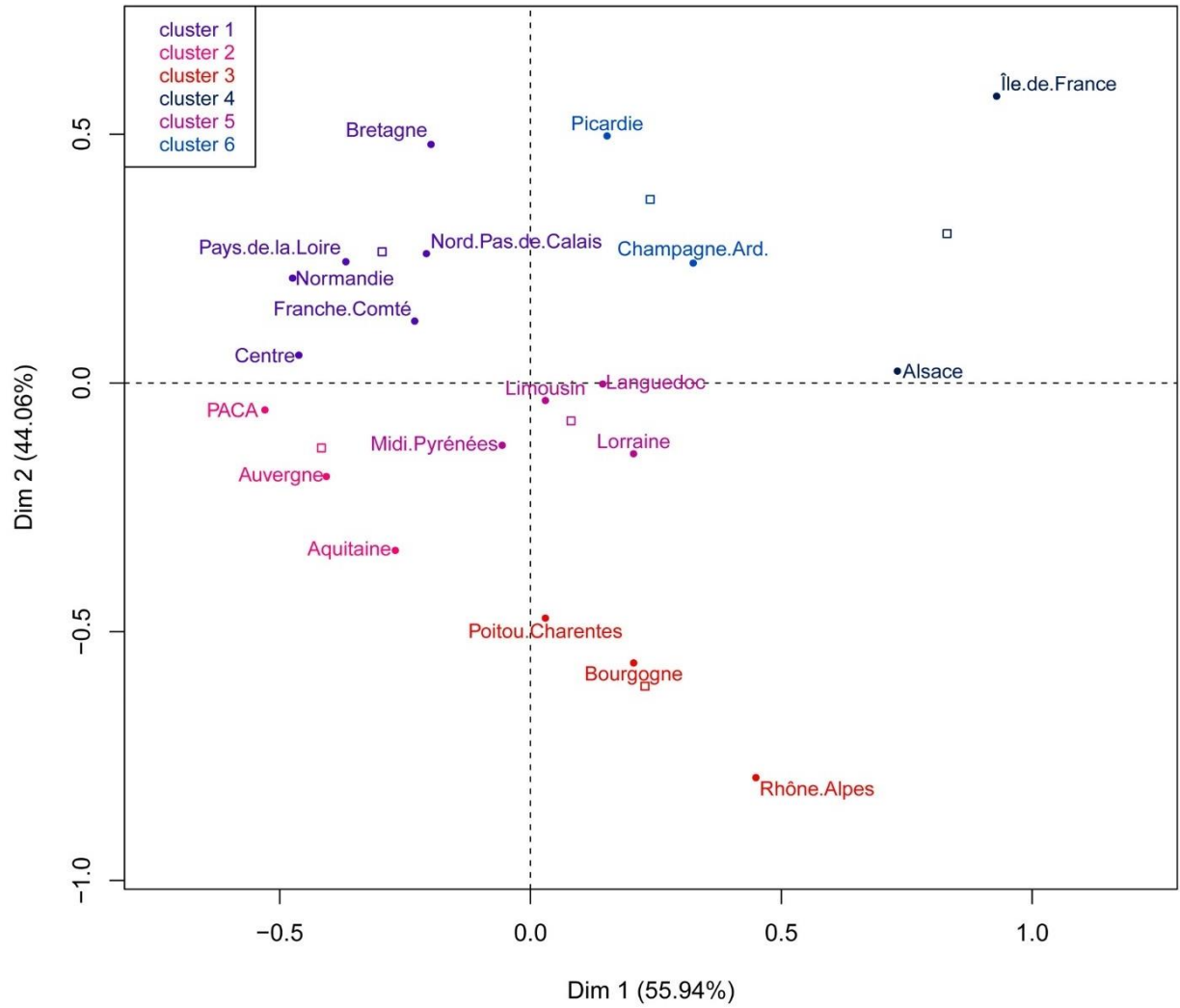

**Fig. 8 :** Corpus des actes originaux de l'actuelle France (Artem), analyse factorielle du nombre d'actes produits par région, puis partitionnement, par demi-siècle (900-1121). Plan factoriel 1-2.

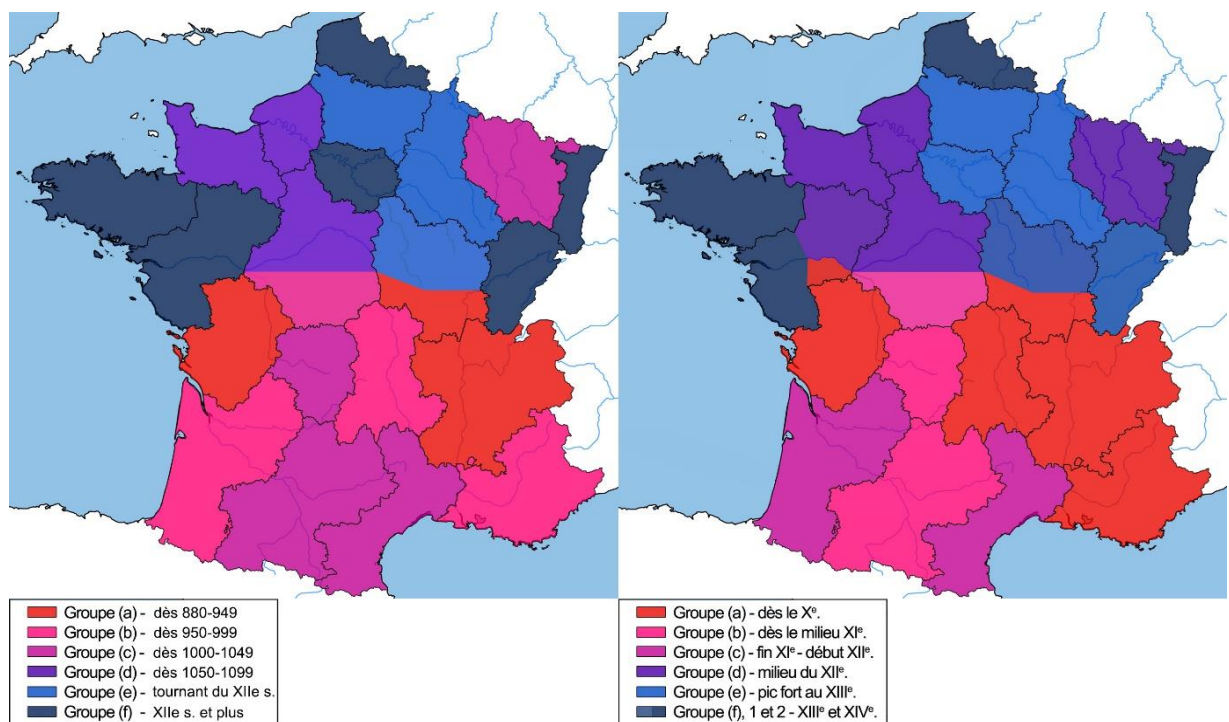

**Fig. 9 :** Corpus des actes originaux de l'actuelle France (Artem) (à gauche, plus de 5 000 actes) et corpus diplomatique incluant les copies (majoritaires) (à droite, plus de 147 000 actes) : répartition chrono-géographique des documents. Projection cartographique à partir d'analyses multivariées.

L'échelle fondamentale de l'analyse des sociétés médiévales est toutefois l'échelle européenne[77]. Un dépouillement aussi large que possible a donc été mené[78], ce qui représente près de 2 000 éditions, soit 520 000 chartes, décomptées par demi-siècle et région. Cette série d'analyse (fig. 10) montre que le pic majoritaire de répartition-production des actes diplomatiques s'est étalé sur environ 4 siècles, mais ne s'est concentré la plupart dans chaque zone que sur un siècle ou un siècle et demi, faisant apparaître des disparités extrêmement fortes.

---

[77] J. LE GOFF, *La civilisation de l'Occident médiéval*, Paris, 1964 ; R. FOSSIER, *Enfance de l'Europe, X$^e$-XII$^e$ siècle*, 2 volumes, Paris, 1982 ; M. MITTERAUER, *Warum Europa? Mittelalterliche Grundlagen eines Sonderwegs*, München, 2003 ; C. WICKHAM, *Medieval Europe*, New Haven, 2016.
[78] Sur Internet, à l'École des chartes, à la Sorbonne, à l'École française de Rome, à l'Université de Münster et dans différents catalogues, en particulier pour l'Espagne : J. Á. CORTAZAR, J.A. MUNITA et L.J. FORTUN (dir.), *Codiphis. Catálogo de colecciones diplomáticas hispano-lusas de época medieval*, 2 volumes, Santander, 1999.

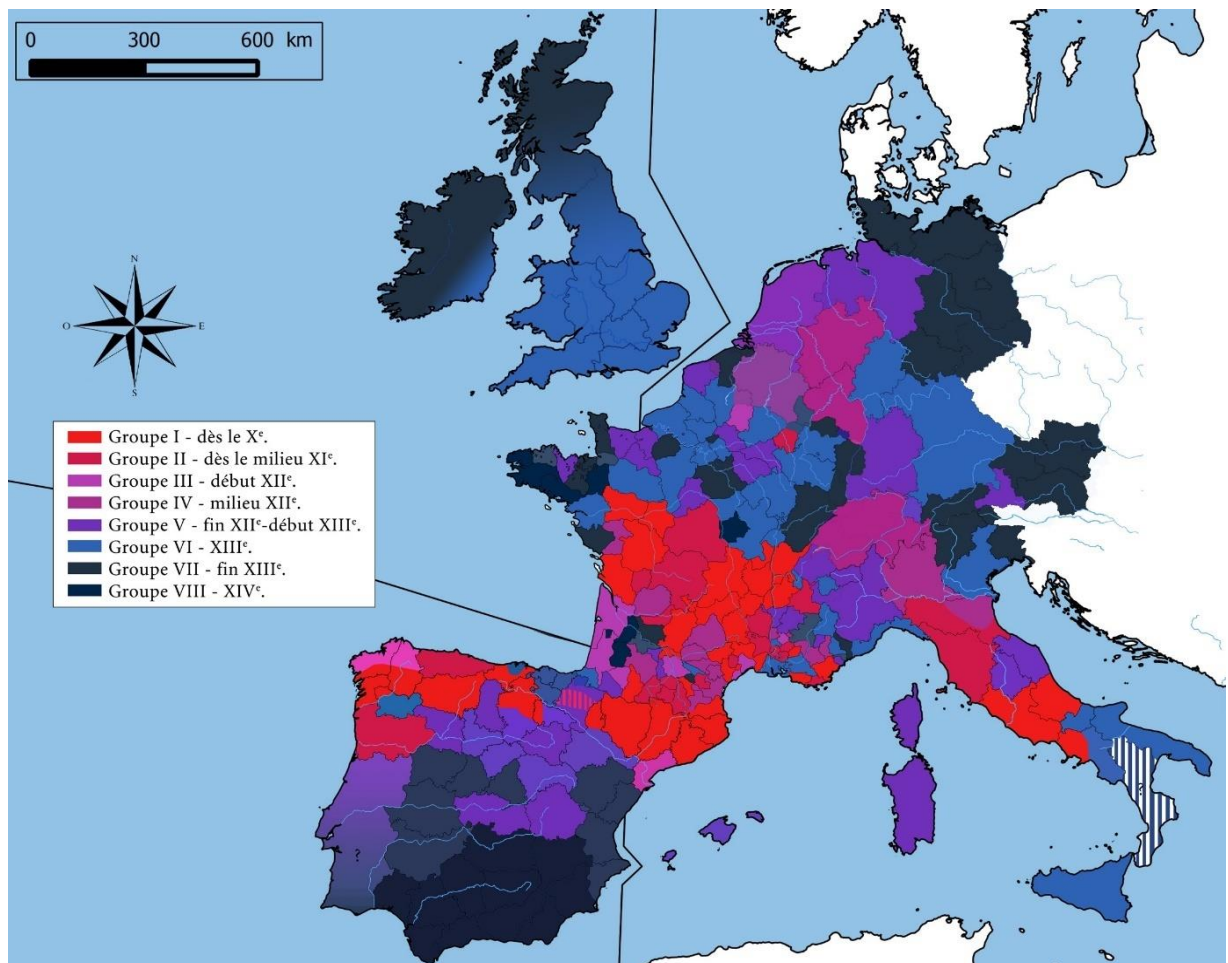

**Fig. 10 :** Corpus diplomatiques européens, répartition chrono-géographique des actes. Projection cartographique à partir d'analyses multivariées portant sur plus de 520 000 actes.

Ainsi, ces visualisations fondées sur le corpus permettent d'observer différents phénomènes difficilement constatables à l'œil nu. C'est parce que le corpus a été considéré comme un tout cohérent, autrement dit possédant un sens social, que les démarches heuristiques ont débuté. Dans le cadre de la thèse, ces méthodes ont entre autres permis de poser l'hypothèse suivante : entre le X[e] et le XIV[e] siècle, la répartition régionale des chartes européenne correspond aux différents moments de la dynamique systémique, très variable d'un espace à l'autre[79].

---

[79] La cartographie des chartes permet ainsi de modéliser le phénomène de densification sociale ayant eu lieu lors de cette période, qualifié d'*incastellamento* par Pierre Toubert, d'encellulement par Robert Fossier puis d'*inecclesiamento* par Michel Lauwers. Le sens de ces concepts est toutefois assez différent. Tandis que l'*incastellamento* se focalisait sur le rôle des châteaux dans l'émergence des communautés, l'encellulement approchait le problème sous angle moins matériel, plus général, voire abstrait – mais toujours pour les X[e]-XII[e] siècles. L'*inecclesiamento* est quant à lui un concept visant une approche de la spatialité médiévale sur le temps long (dès le très haut Moyen Âge), avec comme pôle central les institutions ecclésiales. Cf. P. TOUBERT, *Les structures du Latium médiéval. Le Latium et la Sabine du IX[e] à la fin du XII[e] siècle*, 2 volumes, Rome, 1973 ; R. FOSSIER, *Histoire sociale de l'occident médiéval*, Paris, 1970 ; M. LAUWERS, « De l'incastallamento à l'inecclesiamento. Monachisme et logiques spatiales du féodalisme », in D. IOGNA-PRAT, M. LAUWERS, F. MAZEL et I. ROSE (dir.), *Cluny : les moines et la société au premier âge féodal*, Rennes, 2013, p. 315-338. Voir de même la réflexion sur ces concepts dans J. MORSEL, « La faucille et le goupillon ? Observations sur les rapports entre communauté d'installés et paroisse (XII[e]-XV[e] siècles) », in J. MORSEL (dir.), *Communautés d'habitants médiévales (XII[e]-XV[e] siècles)*, Paris, à paraître.

### 2.3. Le cas des édifices dits « romans »

Il existe évidemment d'autres visuels cartographiques à objectif heuristique. Suite aux analyses précédentes, il était utile de comparer le corpus des chartes à un second. L'objectif était ainsi de déterminer la validité de l'hypothèse posée quant aux liens entre production scripturaire et dynamique sociale. Le corpus des édifices dits « romans » apparaissait comme un point de comparaison intéressant[80]. Il n'existait toutefois pas de telle carte, du moins pas avec un nombre d'édifices suffisant pour une comparaison valable avec les documents écrits (*i.e.* plusieurs milliers). Celle-ci était réputée plus ou moins infaisable, à cause de l'absence de corpus, mais aussi des destructions. Sur quel ensemble était-il possible de se fonder, et quel type de visualisation fallait-il employer ? En dépit des réserves, parfois légitimes, dont elle fait l'objet, la collection de *La nuit des temps* de Zodiaque fournit aujourd'hui encore le matériau le plus structuré en matière d'églises pour ces périodes[81]. Chaque volume contient en effet des cartes localisant les édifices par région : un travail systématique qui confère une certaine homogénéité à la collection[82].

L'ensemble des cartes fournies par Zodiaque a ainsi été géolocalisée, soit une centaine de documents répartis dans 88 volumes. Chaque édifice indiqué a ensuite été pointé manuellement[83]. Même si la couverture obtenue n'est pas encore parfaite, on réalise alors que les volumes donnent la position de plus de 8 600 bâtiments, ce qui est loin d'être négligeable. Pour traiter ce problème de distribution classique, trois méthodes ont été retenues (fig. 11) :

---

[80] Sur ce corpus et les problèmes qu'il pose, voir en dernier lieu : X. BARRAL I ALTET, *Contre l'art roman ? Essai sur un passé réinventé*, Paris, 2006 ; A. HARTMANN-VIRNICH, *Was ist Romanik ? Geschichte, Formen und Technik des romanischen Kirchenbaus*, Darmstadt, 2004 ; J. Nayrolles, *L'invention de l'art roman à l'époque moderne (XVIII$^e$-XIX$^e$ siècles)*, PUR, Rennes, 2005. Pour des raisons évidentes de place, nous ne rentrerons pas ici dans l'historiographie de ces problématiques.

[81] Concernant Zodiaque, voir en premier lieu M. COLLIN, « Les Éditions Zodiaque une aventure de cinquante ans... », in R. CASSANELLI, E. LOPEZ-TELLO (dir.), Benoît et son héritage artistique, Cerf, Paris, p. 425-434 ; C. LESEC, « Zodiaque est une grande chose, maintenant… », in *Revue de l'art*, 157 (2007), p. 39-47 ; C. LESEC, A. SURCHAMP, R. RECHT, P. PLAGNIEUX, O. DELOIGNON (dir.), *Zodiaque. Le monument livre*, Paris, 2012.

[82] Sur le détail des volumes employés, nous nous permettons de renvoyer à N. PERREAUX, « Des structures inconciliables ? Cartographie comparée des chartes et des édifices « romans » (X$^e$-XIII$^e$ siècles) », *op.cit.*

[83] Grâce au logiciel QGIS (https://www.qgis.org/, lien consulté au 26.06.2017).

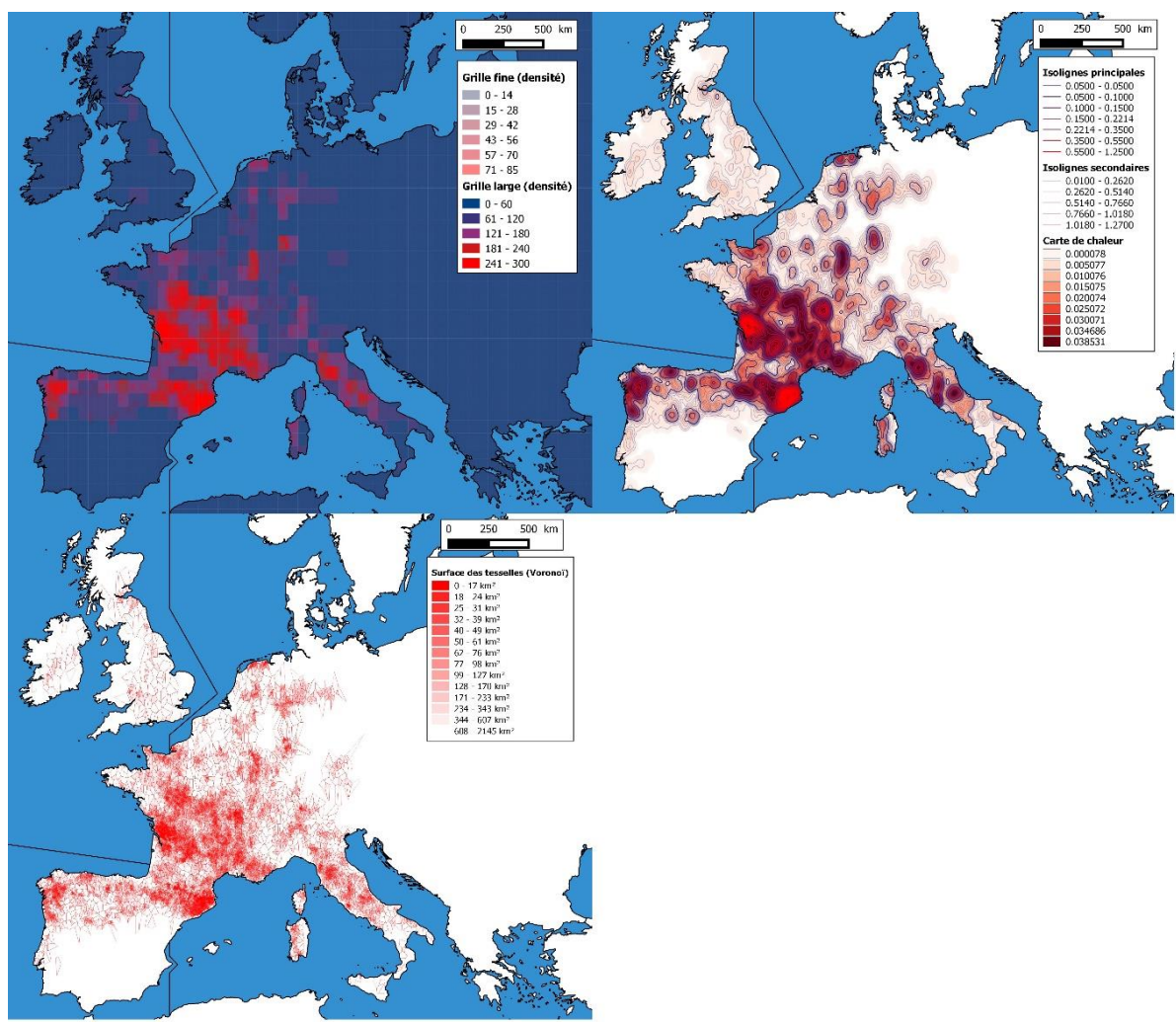

**Fig. 11 :** Europe, densité des 8 600 édifices « romans » recensés, d'après *La nuit des temps* de Zodiaque. Trois méthodes : un double carroyage, en haut à gauche ; une carte de chaleur associée à une estimation de densité de noyau, en haut à droite ; polygones de Voronoï, en bas à gauche.

Or, ce que nous observons ici, tant à l'échelle globale que dans le détail, c'est que l'adéquation entre les zones possédant un pic d'actes pour les X$^e$-XI$^e$, voire XII$^e$ siècles, et les zones possédant une forte densité d'édifices dits « romans » est presque totale. Cette cohérence permet d'énoncer, au-delà des destructions[84], que peu ou pas d'édifices dits romans furent édifiés dans certains zones, et que le nombre de bâtiments pour cette période correspond au degré variable de dynamique sociale locale[85].

En définitive, les deux méthodes de visualisation, fondées sur l'exploration d'un corpus par la cartographie numérique, d'une part *via* la projection d'analyses factorielles (pour les chartes), d'autre part grâce à des analyses de densité (pour les édifices), paraissent avoir fonctionnées de façon heuristique. Elles ont permis de révéler des structures jusqu'alors invisibles à l'œil nu, et de proposer un modèle chrono-géographique de la dynamique européenne entre les X$^e$-XIV$^e$ siècles, qui n'avait pas été envisagé antérieurement. Dans les deux cas, la formalisation est restée extrêmement simple, mais la logique de l'analyse historienne a été « inversée », puisque les visuels ont servi de base à celle-ci et n'ont pas été employés à titre de « témoignages ».

---

[84] L. REAU, *Histoire du vandalisme. Les monuments détruits de l'art français*, édition augmentée par Michel Fleury et Guy-Michel Leproux, Paris, 1994.
[85] Sur la question de la « monumentalisation », voir É. ZADORA-RIO (dir.), *Des paroisses de Touraine aux communes d'Indre-et-Loire. La formation des territoires*, Tours, 2008.

## 3. Représenter le texte

Quels autres types de visualisations pourraient-ils rendre service à l'historien ? Sans prétendre explorer toutes les voies envisageables (qui restent d'ailleurs largement à découvrir), quelques possibilités complémentaires seront présentées, cette fois fondées sur le texte lui-même. Lors de l'introduction et de la première partie, la question de la tension entre le visuel et le langage, puis entre une analyse fondée sur des dispositifs « visuels heuristiques » et une analyse logocentrée, fondée sur l'exemplarité, a été évoquée. En dépit des difficultés spécifiques qu'il pose, le texte reste souvent au cœur du travail de l'historien, y compris en ce qui concerne les chartes, qui sont avant tout envisagées en tant que document textuel[86]. Ce qui pose bien évidemment l'épineuse question du sens des mots qu'il contient[87]. Dans quelle mesure la visualisation pourrait-elle aider l'historien à proposer de nouvelles hypothèses, et par-là de nouveaux modèles ?

### 3.1. Mots, formules, formulaires

Une des questions fréquentes dans la médiévistique des dernières années est celle de la circulation et des influences scripturaires[88]. La plupart du temps toutefois, la recherche de circulations ou de « contaminations » lexicographiques s'effectue sur la base de quelques pièces documentaires et selon le paradigme de l'exemplarité. Telle formule, présente dans tel texte, se retrouve dans un autre établissement d'un autre espace, etc. Est-il possible d'aller plus loin, encore une fois grâce à la production de figures à vocation heuristique ?

---

[86] Avec néanmoins une inflexion profonde depuis quelques années vers l'étude de la matérialité documentaire.
[87] L'article d'Alain Guerreau consacré au terme *vinea* est devenu assez emblématique de la question, en médiévistique. La problématique est cependant plus ancienne, et renvoie largement à la linguistique structurale. A. GUERREAU, « Vinea », in M. GOULLET et M. PARISSE (dir.), *Les historiens et le latin médiéval*, Paris, 2001, p. 67-73.
[88] Concernant les « formules », voir en dernier lieu E. LOUVIOT (dir.), *La formule au Moyen Âge*, Turnhout, 2012 ; I. DRAELANTS et C. BALOUZAT-LOUBET (dir.), *La formule au Moyen Âge II*, Turnhout, 2015. En diplomatique, voir M. ZIMMERMANN (dir.), *Auctor et Auctoritas. Invention et conformisme dans l'écriture médiévale*, Paris, 2001 ; M. ZIMMERMANN., *Écrire et lire en Catalogne (IX$^e$-XII$^e$ siècle)*, 2 volumes, Mardrid, 2003 ; M. TOCK, « L'apport des bases de données de chartes pour la recherche des mots et des formules », in G. VOGELER (dir.), *Digitale Diplomatik : neue Technologien in der historischen Arbeit mit Urkunden*, Cologne, 2009, p. 283-293 ; I. ROSE, « Judas, Dathan, Abiron, Simon et les autres. Les figures bibliques-repoussoirs dans les clauses comminatoires des actes originaux français », in *Archiv für Diplomatik*, 62 (2016), p. 59-106.

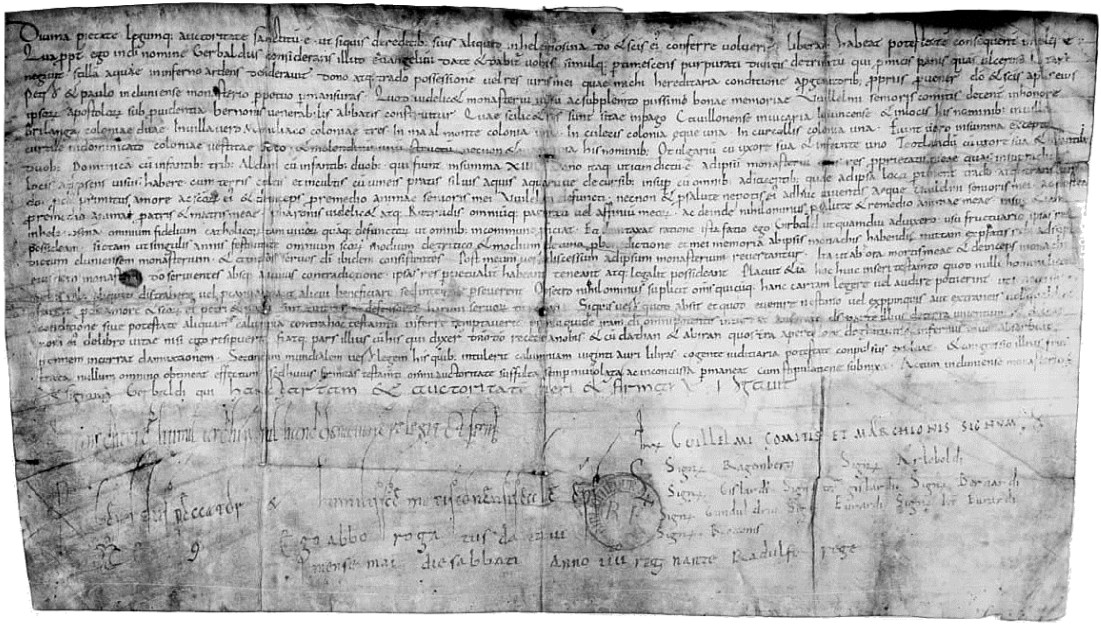

**Fig. 12 :** Charte clunisienne BB n° 269 (926). Cliché Artem, Nancy.

L'acte clunisien présenté ici (fig. 12), daté de 926, montre combien il est délicat de tracer une formule et ses multiples variations dans un ensemble documentaire[89]. Son préambule, qui évoque la Parabole de Lazare et du riche, est par la suite repris à trois reprises, en février et mars 941, puis en 943-944[90]. Un examen du corpus permet encore de constater que cette charte a servi de modèle à de nombreuses autres, qui la reprennent différents passages. La charte débute ainsi par la formule : « *Divina pietate legumque auctoritate [...]* », qui est reprise 33 fois dans la documentation clunisienne[91]. Ne pouvant étudier la diffusion de chaque partie de cet acte – les ramifications couvrant plusieurs centaines de chartes –, comment est-il possible de le situer au sein de la production diplomatique clunisienne ? Grâce au programme Text-to-CSV[92], 25 000 tri-formes[93] ont été extraits de l'ensemble des actes datés de l'abbaye, soit 4 668 documents[94]. Différentes méthodes de réduction de dimentionnalité au tableau rendent alors possible la visualisation de la dynamique scripturaire du fond clunisien (fig. 13). Et donc la localisation de l'acte de 926 et de sa descendance dans la généalogie textuelle de Cluny.

---

[89] A. BERNARD et A. BRUEL (éd.), *Recueil des chartes de l'abbaye de Cluny*, 6 volumes, Paris, 1876-1903 (*Collection de documents inédits sur l'histoire de France. Première série, Histoire politique*), désormais BB. Il s'agit ici de la charte *BB* n° 269.
[90] BB, n° 524 (février 941), 527 (mars 941) et 649 (943-944).
[91] Par exemple BB, n° n° 345 (927-942), n° 392 (931), n° 649 (943-944) et n° 807 (951).
[92] Dont le fonctionnement schématique est décrit dans N. PERREAUX, « L'écriture du monde (I). Les chartes et les édifices comme vecteurs de la dynamique sociale dans l'Europe médiévale (VII$^e$-milieu du XIV$^e$ siècle) », *op.cit.* ; ID., « L'écriture du monde (II). L'écriture comme facteur de régionalisation et de spiritualisation du *mundus* : études lexicales et sémantiques », *op.cit.*
[93] C'est-à-dire des groupes de trois mots.
[94] On obtient alors un tableau de 116 millions de cases, soit autant de relations textuelles possibles.

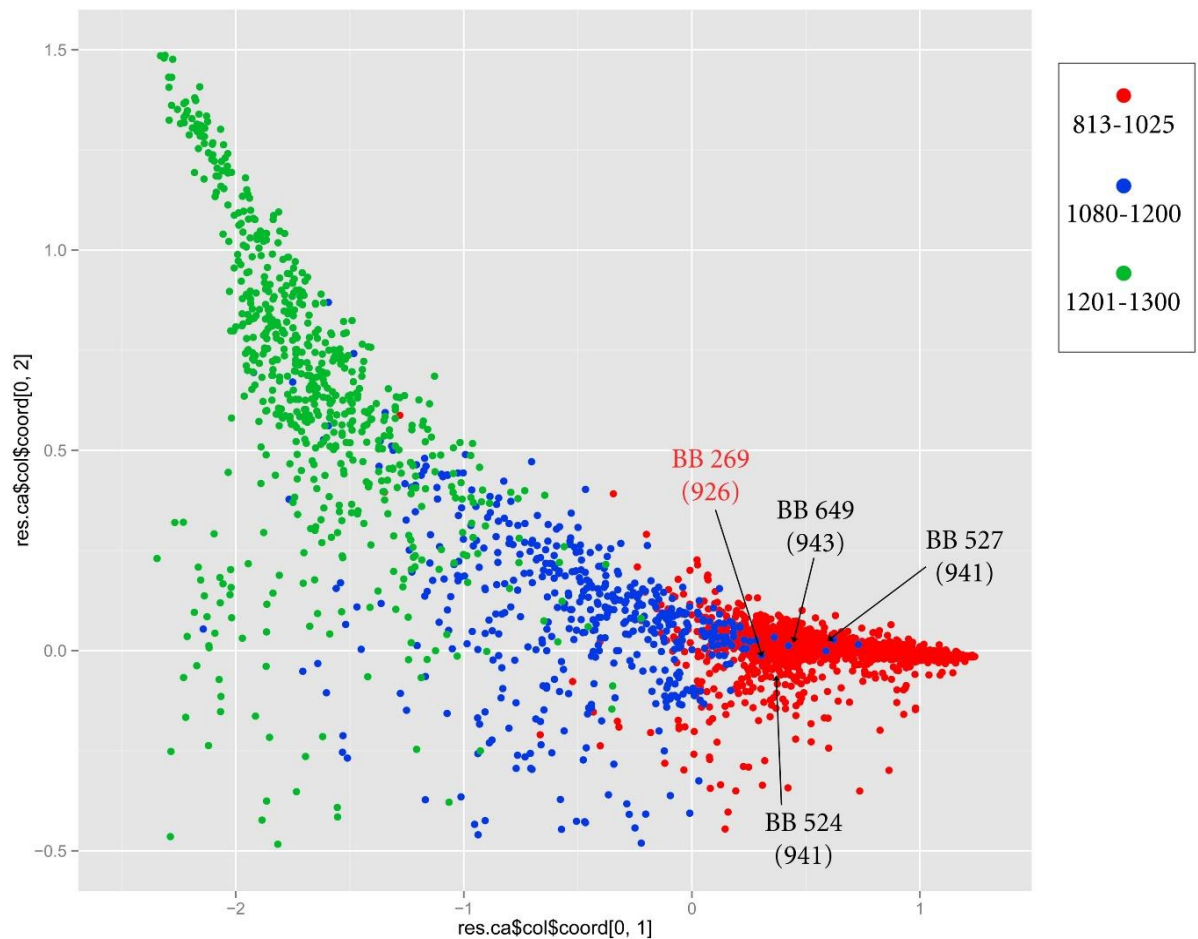

**Fig. 13 :** Cluny, évolution de la totalité du lexique dans l'édition de Bernard et Bruel (25 000 tri-formes), par analyse factorielle. Chaque point correspond à un acte diplomatique.

Ces méthodes visuelles pourraient-elle être généralisées ? Des expériences similaires ont par la suite été menées sur l'ensemble des 140 000 chartes actuellement disponibles. Menées sur les 500 lemmes les plus fréquents du lexique diplomatique (fig. 14), puis sur 16 000 bi-formes, les analyses font ressortir des répartitions correspondant aux grandes aires de l'Europe contemporaine, placées en couleur sur les analyses factorielles. Un examen permet même de remarquer qu'une rotation à 90° dans le sens inverse des aiguilles d'une montre de la carte factorielle fait apparaître un graphique ressemblant très fortement à une carte géographique de l'Europe. Des analyses complémentaires permettent alors de faire ressortir la progressive unification du vocabulaire des actes, dans un contexte d'émergence européenne (fig. 15).

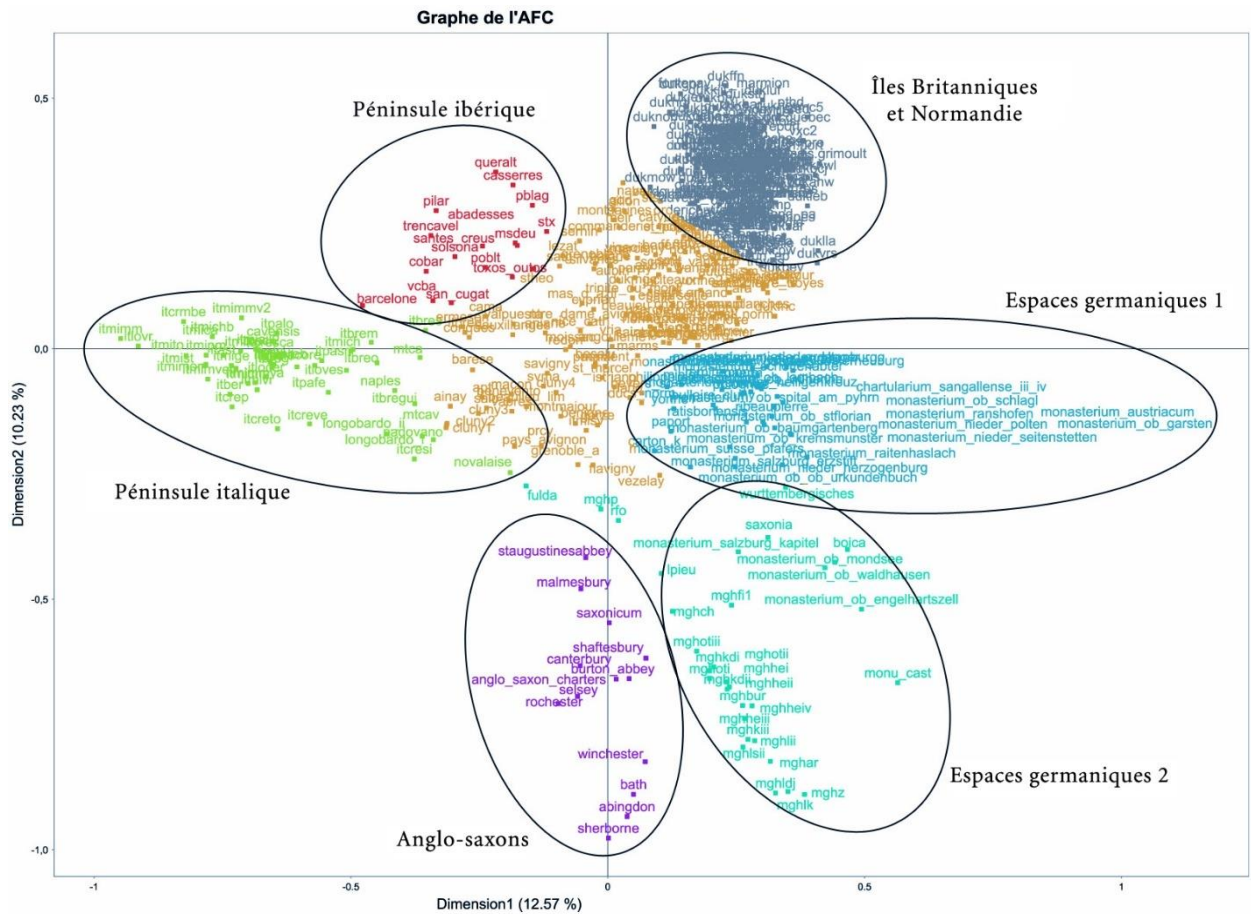

**Fig. 14 :** Corpus diplomatique européen (CEMA), distribution du lexique. Analyse factorielle du tableau de contingence automatique des cinq cents lemmes les plus fréquents au sein du corpus. Plan factoriel 1-2.

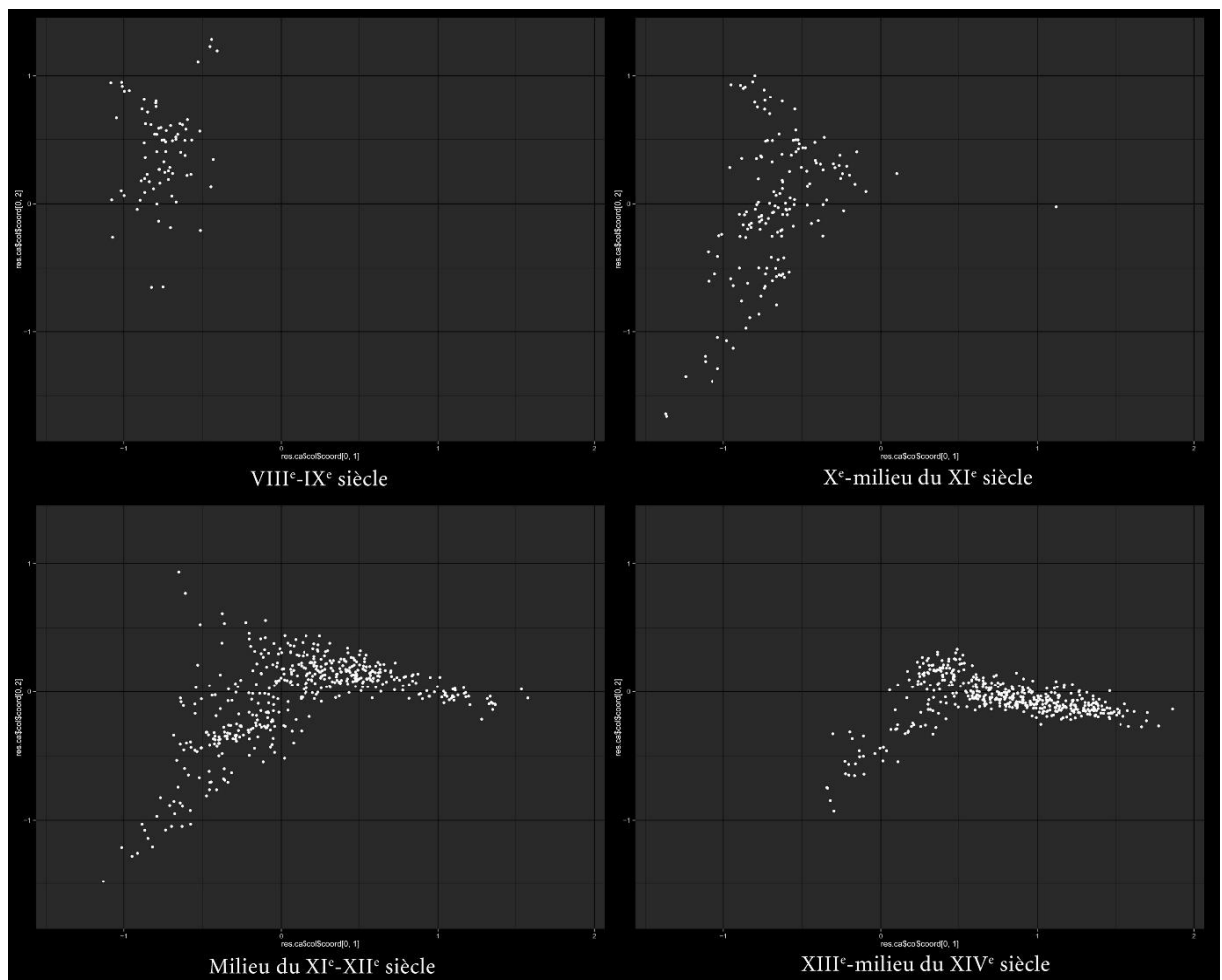

**Fig. 15 :** Corpus diplomatique européen (CEMA), évolution générale du lexique (16 000 bi-formes) des actes diplomatiques dans le corpus. Chaque point représente un paquet documentaire (= un ensemble documentaire, divisé par période et nettoyé). Analyse chronologique seule.

### 3.2. Distinguer les « auteurs »

Un autre domaine où ces méthodes « stylométriques », associant calcul numérique et visualisation, pourraient produire des résultats positifs est celui de l'attribution d'auteur (*Authorship attribution*). Constituer un corpus autour d'un unique producteur médiéval est bien souvent délicat[95]. Or, la fouille de données, en donnant une probabilité d'attribution d'un groupe de textes à un même personnage, permet de mieux circonscrire ce type de corpus. Prenons un exemple, auquel nous consacrons actuellement une enquête : la vie de saint Jean de Réome, fondateur de l'abbaye du même nom en Bourgogne[96] au milieu du VIe siècle, attribuée à Jonas

---

[95] Trois exemples parmi tant d'autres, volontairement pris à plusieurs siècles d'écart : saint Augustin, Haymon d'Auxerre, Honorius Augustonensis.

[96] Cette enquête s'inscrit dans le cadre d'un travail plus vaste, mené par deux groupes de travail, sur et autour du monastère. Ici, nous pensons en particulier à un volume autour du manuscrit n° 1 de la Bibliothèque municipale de Semur-en-Auxois, en cours d'élaboration sous la direction d'Eliana Magnani et Daniel Russo, auquel nous collaborons aux côtés d'Eduardo Aubert. Trois riches bilans d'étape ont été publiés dans : E. MAGNANI et D. RUSSO, « Le manuscrit 1 de la Bibliothèque municipale de Semur-en-Auxois, provenant de l'abbaye de Saint-Jean de Réome (Moutiers-Saint-Jean) : programme pédagogique de recherche », in *Bulletin du Centre médiéval d'Auxerre*, 12 (2008) (en ligne : https://cem.revues.org/7212) ; E. AUBERT, E. MAGNANI et D. RUSSO, « Histoire du manuscrit médiéval, transmission des textes, des chants, compositions peintes et décors. Le manuscrit 1 de la Bibliothèque municipale de Semur-en-Auxois au miroir de ses manuscrits « voisins » », in *Bulletin du Centre*

de Bobbio[97]. Moine célèbre de la première moitié du VII[e] siècle, ce dernier a composé une importante série de textes hagiographiques : vies de abbés Attala (BHL 742) et Bertholphe (BHL 1311), de Colomban (BHL 1898), d'Eustache de Luxeuil (BHL 2773) et de Fare (Burgundofara, BHL 1487)[98].

Le dossier hagiographie de saint Jean de Réome correspond aux numéros BHL 4424 à 4431[99]. Une édition de la *vita* (dite n°1, BHL 4424), donnée par Bruno Krusch, est disponible dans la collection des MGH[100]. Le fameux éditeur avait considéré que le manuscrit conservant la version la plus fiable de la *vita* était un manuscrit du XIV[e] siècle, aujourd'hui conservé à la Bibliothèque nationale (BnF, ms. lat. 5306)[101]. En 2017, deux nouvelles éditions sont en cours de réalisation, d'une part pour la BHL 4424 et d'autre part pour la BHL 4426[102]. Pour cette dernière, le plus ancien témoin est un manuscrit provenant de l'abbaye de Réome, daté du tournant des X[e]-XI[e] siècles, aujourd'hui conservé à la Bibliothèque municipale de Semur (Bibliothèque municipale de Semur, ms. 1). C'est sur l'édition de ce document que nous réalisons les analyses qui suivent[103].

Une formalisation identique aux précédentes a tout d'abord été appliquée au corpus des documents hagiographiques attribués à Jonas (dont la BHL 4426 donc), puis à ce corpus additionné de quatre *vitae* clunisiennes[104].

---

*médiéval d'Auxerre*, 13 (2009) (en ligne : https://cem.revues.org/11055) ; E. AUBERT, E. MAGNANI et D. RUSSO, « Le manuscrit 1 de Semur-en-Auxois », in *Bulletin du Centre médiéval d'Auxerre*, 14 (2010) (en ligne : https://cem.revues.org/11561). Voir en outre E. MAGNANI, « Hagiographie et diplomatique dans le monachisme réformé en Bourgogne au miroir du manuscrit 1 se Semur-en-Auxois », in M.-C. ISAÏA et T. GARNIER (dir.), *Normes et hagiographie dans l'Occident latin (VI[e]-XVI[e] siècles)*, Turnhout, 2014, p. 183-195.

[97] Les travaux sur Jonas de Bobbio sont nombreux depuis quelques années, en particulier à partir de la vie de saint Colomban. Voir en dernier lieu A. O'HARA, *Jonas of Bobbio and Vita Columbani: sanctity and community in the seventh century*, St Andrews, 2009 ; B. JUDIC, « Jonas of Bobbio », in A. VAUCHEZ, R.B. DOBSON et M. LAPIDGE (dir.), *Encyclopedia of the Middle Ages*, Chicago, 2000, tome 1, p. 783-784 ; I. WOOD, « Jonas of Bobbio, The Abbots of Bobbio from the Life of St. Columbanus », in T. HEAD (dir.), *Medieval hagiography. An anthology*, New York, 2000, p. 111-135 ; I. PAGANI, « Jonas von Bobbio (Ionas von Susa), italienischer Mönch und Hagiograph († nach 659) », in *Lexikon des Mittelalters*, München – Zürich, 1980-1998, tome 5, p. 624-625. Concernant la *vita Johannis*, voir A. DIEM, « The rule of an "Iro-Egyptian" monk in Gaul. Jonas' *Vita Iohannis* and the construction of a monastic identity », in *Revue Mabillon*, n.s., 19 (2008), p. 5-50.

[98] JONAS DE BOBBIO, *Vita Sancti Attalae abbatis Bobbiensis secundi*, *Patrologie latine* (=PL) vol. 87, col. 1055-1062 ; ID., *Vita Sancti Bertulfi abbatis Bobbiensis tertii*, vol. 87, col. 1062-1070 ; ID., *Vita Sancti Columbani abbatis*, PL, vol. 87, col. 1011-1046 ; ID., *Vita Sancti Eustasii abbatis Luxoviensis secundi*, PL, vol. 87, col. 1046-1056a ; ID., *Vita Sanctae Burgundofarae abbatissae Eboriacensis primae*, PL, vol. 87, col. 1070-1084b.

[99] On trouvera les références aux éditions dans E. AUBERT, E. MAGNANI et D. RUSSO, « Histoire du manuscrit médiéval, transmission des textes, des chants, compositions peintes et décors. Le manuscrit 1 de la Bibliothèque municipale de Semur-en-Auxois au miroir de ses manuscrits « voisins » », *op.cit.*, note 5.

[100] JONAS DE BOBBIO (?), « *Vita Ihoannis abbatis Reomaensis auctore Iona* », in B. KRUSCH (éd.), *MGH, Scriptores rerum merovingicarum, III*, Hanovre, 1896, p. 505-517 et ID., *MGH, SS rer. Germ. in usum scholarum*, 1905, p. 321-344. Il existe une traduction récente de ce texte : A. O'HARA et I. WOOD, *Jonas of Bobbio. Life of Columbanus, Life of John of Réomé, and Life of Vedast*, Liverpool, 2017 (Translated Texts for Historians 64).

[101] Paris, Bibliothèque nationale de France, ms. lat. 5306, fol. 65-67. Krusch décrit ainsi cette version : « *formaem agnae, […] textum Vitae plenum continet, nisi quod prologi cum laterculo capitum desiderantur* », dans JONAS DE BOBBIO (?), « *Vita Ihoannis abbatis Reomaensis auctore Iona* », *op.cit.*, p. 503.

[102] D'une part pour la BHL 4426 par Eduardo Aubert, dans le cadre d'un volume placé sous la direction d'Eliana Magnani et Daniel Russo, dont nous participons à l'élaboration. D'autre part, pour la *vita* BHL 4424, par Alain Dubreucq.

[103] Il manque toutefois dans le codex les derniers feuillets du livret hagiographique, qui peuvent être complétés grâce à la copie du texte contenue dans le manuscrit BAV, Reg. lat. 493 (XI[e] siècle, Auxerre).

[104] Soit Maïeul par Odilon (ODILON, *Maiolus, abbas Cluniacensis, Vita auctore Odilone abbate Cluniacensis*, PL, vol. 142, col. 943-926b (BHL 5182)) ; Guillaume de Volpiano par Raoul Glaber (RAOUL GLABER, *Willelmus abbas Sancti Benigni Divionensis, Vita auctore Rodulfo Glabro*, in V. GAZEAU et M. GOULLET, *Guillaume de Volpiano, un réformateur en son temps (962-1031). Vita domni Wilhelmi de Raoul Glaber (Bibliographie, édition et traduction)*, Caen, 2008, ainsi que PL, vol. 142, col. 698-720b (BHL 8907)) ; Odon par Jean de Salerne (JEAN DE SALERNE, *Joannes Italus Cluniacensis, Vita Sancti Odonis Abbatis Cluniacensis II*, PL, vol. 133, col. 43-86

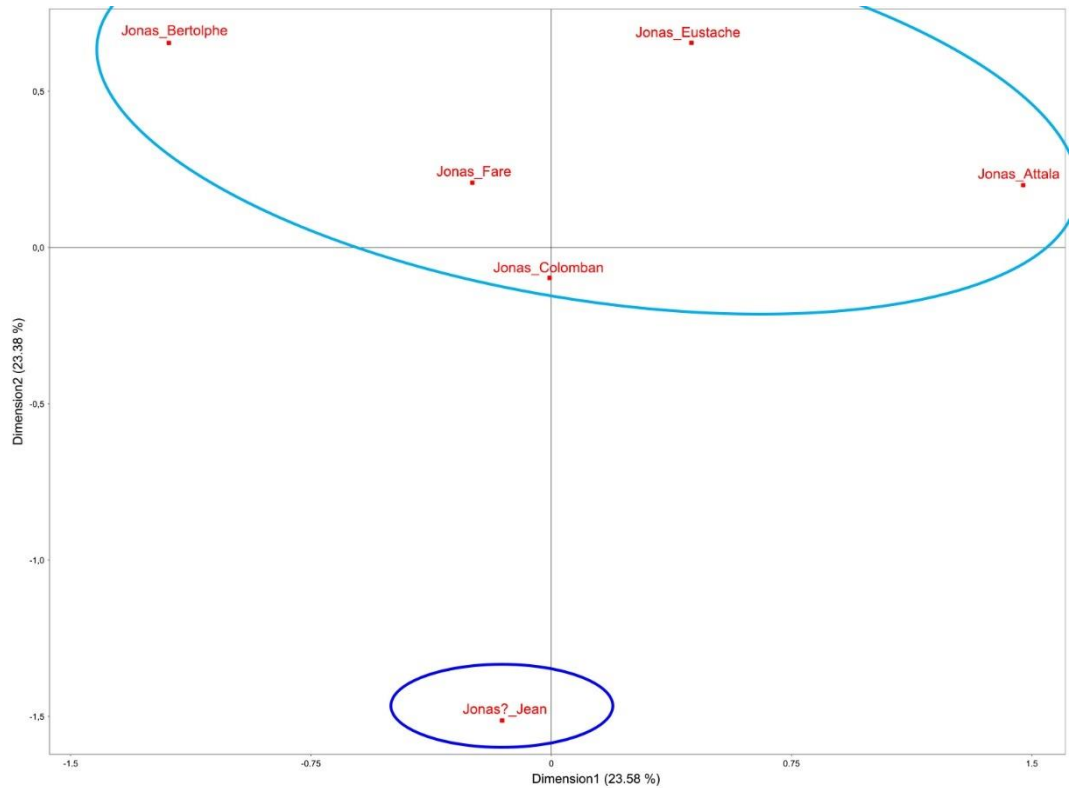

**Fig. 16 :** Vies attribuées à Jonas de Bobbio. Distance lexicale entre les textes, via un traitement par analyse factorielle du lexique (plan factoriel 1-2). On constate que la vie de Jean de Réome (BHL 4426) s'oppose à toutes les autres sur l'axe 2.

---

(BHL 6292)) ; Géraud d'Aurillac par Odon (ODON, *De Vita Sancti Geraldi Aureliacensis Comitis Libri Quatuor*, in A.-M. BULTOT-VERLEYSEN (éd.), *Odon de Cluny. Vita sancti Geraldi Auriliacensis. Édition critique, traduction française, introduction et commentaires*, Bruxelles, 2009 (*Subsidia hagiographica*, 89), et aussi PL, vol. 133, 639b–710c (BHL 3411)). Il s'agit d'un ensemble volontairement varié, typologiquement et chronologiquement.

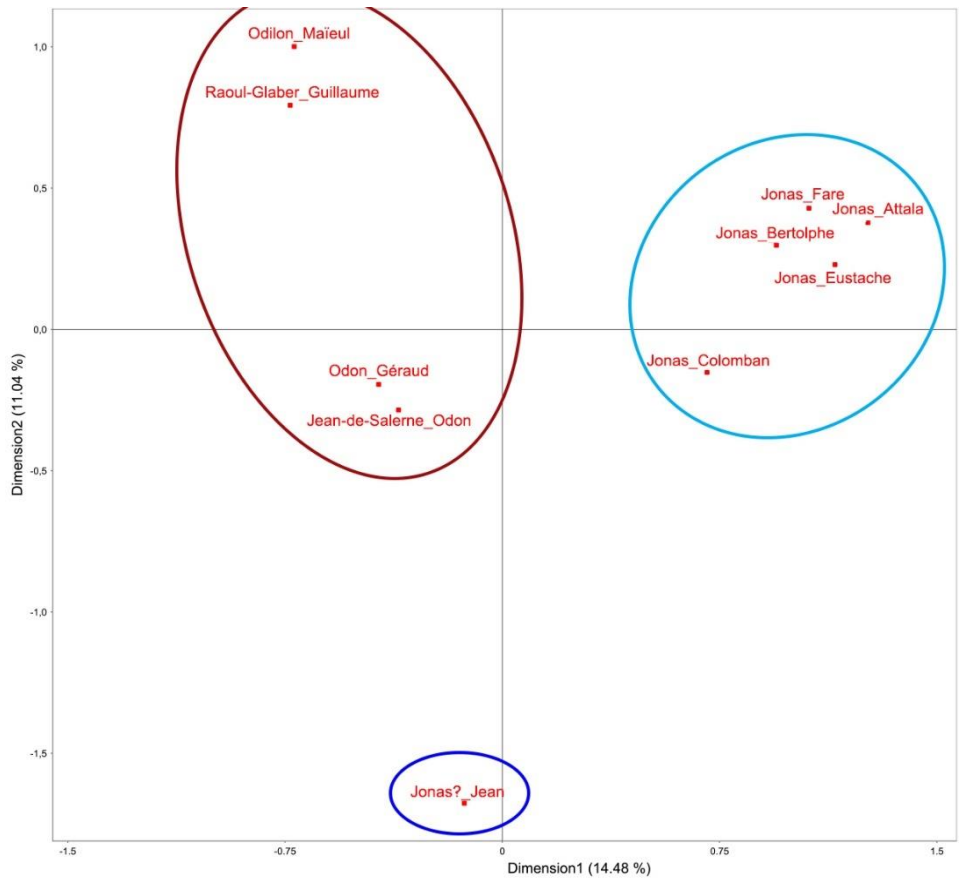

**Fig. 17 :** Corpus de *vitae* : vies attribuées à Jonas de Bobbio et vies liées à Cluny. Distance lexicale entre les textes, via un traitement par analyse factorielle du lexique (plan factoriel 1-2). Les vies attribuées à Jonas se regroupent automatiquement, sauf celle de saint Jean de Réome (BHL 4426). Les vies liées à Cluny se regroupent de la même façon.

Dans toutes les configurations d'analyse, la vie de Jean de Réome BHL 4426, issue du ms. 1 de Semur, se place automatiquement à distance de toutes les autres attribuées à l'auteur. Sous réserves d'analyses complémentaires, ces deux expériences laissent ainsi penser 1) soit que la *vita* de saint de Réome BHL 4426 contenue dans le ms. 1 de Semur n'est pas de Jonas de Bobbio ; 2) soit qu'il s'agit d'une version très remaniée d'un texte de Jonas[105]. Il ne s'agit certes que d'un exemple, mais on peut imaginer que de telles procédures soient systématisées sur les corpus numériques du Moyen Âge latin, permettant de reconsidérer des ensembles textuels parfois édités depuis plus d'un siècle.

### 3.3. Sémantique historique et visualisation

Le dernier groupe de méthodes articulant visualisation et approche heuristique qu'il semblait intéressant de présenter relève de ce que l'on nomme la sémantique historique[106]. Les médiévistes, tout comme les antiquisants, sont face à un problème crucial : celui de l'altérité des sociétés qu'ils étudient. Il est de fait quasi-impossible de reconstruire la dynamique des groupes humains sans comprendre leur vision du monde, leurs normes, leurs objectifs. Dans cette perspective, la modélisation du sens par des méthodes de visualisation apparaît comme un progrès décisif pour la reconstruction abstraite de ces sociétés. De simples graphes fréquentiels peuvent par exemple nous apprendre beaucoup sur le sens d'un terme et, évidemment, sur sa diffusion dans le temps et dans l'espace.

---

[105] A ce titre, il s'agirait désormais de comparer les *vitae* BHL 4424 et 4426, afin de mesurer leurs points communs et divergences.
[106] Voir en dernier lieu le travail minutieux de E. MÜLLER et F. SCHMIEDER, *Begriffsgeschichte und historische Semantik - Ein kritisches Kompendium*, Berlin, 2016.

Toutefois, une étape sans doute plus avancée de la visualisation est évidemment la reconstruction du « champ » ou plutôt des réseaux de cooccurrents autour d'un mot ou d'un groupe de mots. Partant de listes de cooccurrents – c'est-à-dire de termes apparaissant dans le contexte plus ou moins immédiat d'un mot pivot – il est possible de modéliser l'évolution de ces termes et donc d'analyser, à partir de graphiques, l'évolution du sens d'un terme. On peut aussi approcher la question du sens par la reconstruction desdits réseaux sémantiques. Il existe de nombreuses façons de générer ce type de graphes : il y a donc là un champ de recherche assurément neuf et prometteur. On peut enfin aller plus loin et reconstruire l'ensemble du contexte autour d'un groupe de terme, pour comprendre comment ceux-ci forment une structure :

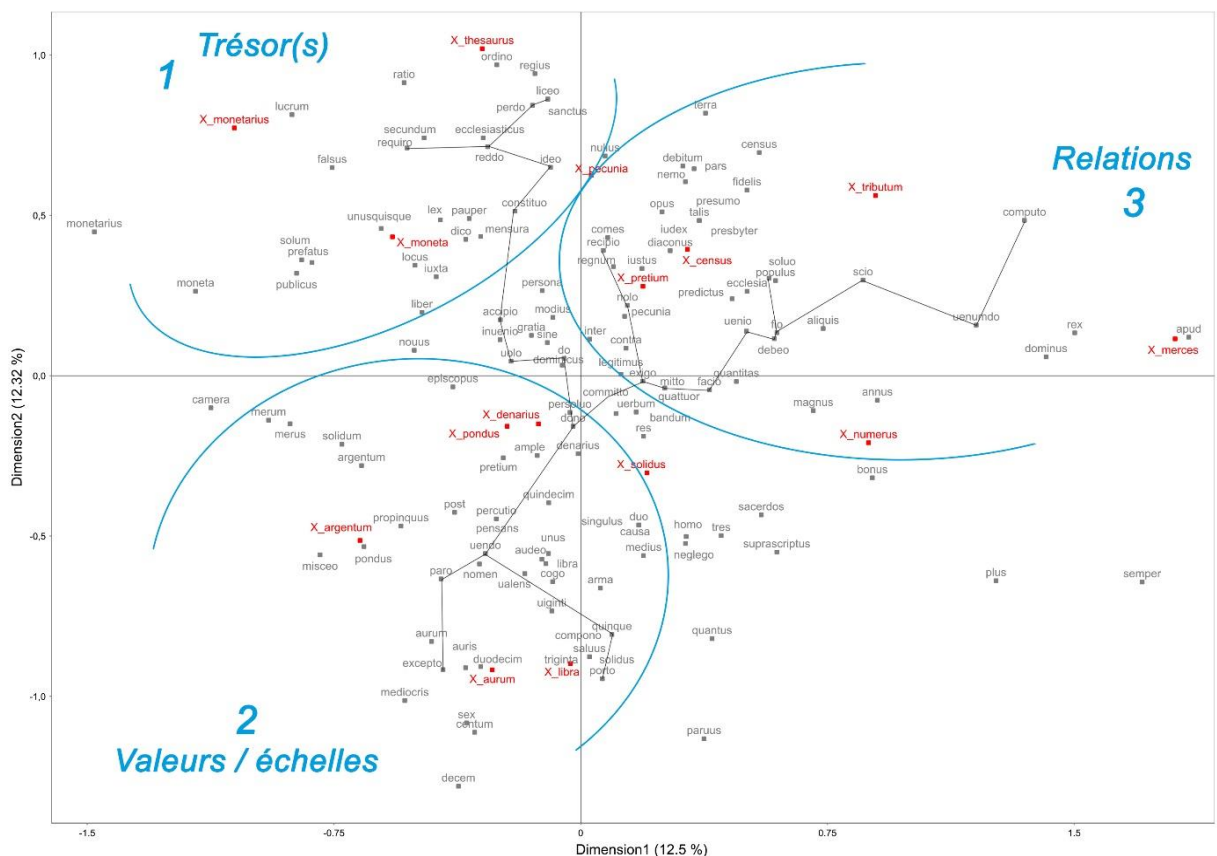

**Fig. 18 :** Un exemple d'analyse sémantique : les relations entre les termes relatifs aux monnaies médiévales, dans la collection des Capitulaires des MGH. A partir d'une liste de 15 termes, on obtient un espace lexical, dont il s'agit de saisir l'organisation interne et la logique. En rouge : les mots sélectionnés ; en gris : les cooccurrences. Plus les termes sont proches sur le graphique, plus ils apparaissent dans des espaces sémantiques communs. Ici, le champ se divise en trois ensembles, définissant trois domaines liées aux monnaies : la thésaurisation, la définition de valeurs et d'échelles, la production de relations sociales[107].

---

[107] Ce graphique a été initialement présenté lors d'une communication avec Ludolf Kuchenbuch, en septembre 2016, à l'occasion d'une conférence à Göteborg (Suède) « Words Don't Come Easy Questions about Historical Semantics in Medievalist and Early Modern Research » (org. Cordelia Heß, Wojtek Jezierski et Gadi Algazi). Son intitulé était : « Modelling the Early Middle Ages? Micro- and macro-semantics Approaches of Lexical Fields: A Comparison ».

**Éléments conclusifs**

Les possibilités de visualisation à partir des documents médiévaux sont immenses, et ceci d'autant plus que les historiens ont déjà numérisés une part conséquente de l'héritage culturel, essentiellement de l'Antiquité aux X$^e$-XII$^e$ siècles. Tandis qu'à l'heure actuelle, les graphes d'historiens (même employés en tant que « témoignages ») restent rares, et que ceux conçus dans une perspective heuristique sont presque inexistants, la discipline pourrait probablement largement profiter de ces dispositifs visuels. Quelques exemples ponctuels ont été présentés, mais il serait évidemment souhaitable de diversifier les techniques : analyse de réseaux, production de « stemmas 2.0 », chronologies interactives, etc.

Un des progrès les plus importants à attendre de ces méthodes visuelles est aussi une meilleure définition des modèles de l'essor de l'Europe médiéval, très variable d'un espace à l'autre. Jusqu'ici, de telles reconstructions se sont heurtées à la complexité des textes et des objets légués par ces sociétés, dont la compréhension implique la prise en compte simultanée d'un nombre faramineux de facteurs : agriculture, organisation spatio-temporelle, parenté, système monétaire, marchés, rapport au monde, etc. En permettant la création de dispositifs visuels faisant apparaître les relations dynamiques entre ces différents pans de la société, ces méthodes pourraient bien révolutionner notre compréhension des sociétés anciennes à plus ou moins court terme.

Enfin, ces visuels amènent à nous interroger sur la nature même et le fonctionnement des dispositifs visuels à vocation heuristique, particulièrement en histoire. En 1966, le volume d'*Introduction au monde des symboles* de la collection Zodiaque proposait des rapprochements originaux entre des objets médiévaux et des photographiques issues des sciences de la vie et de la terre[108].

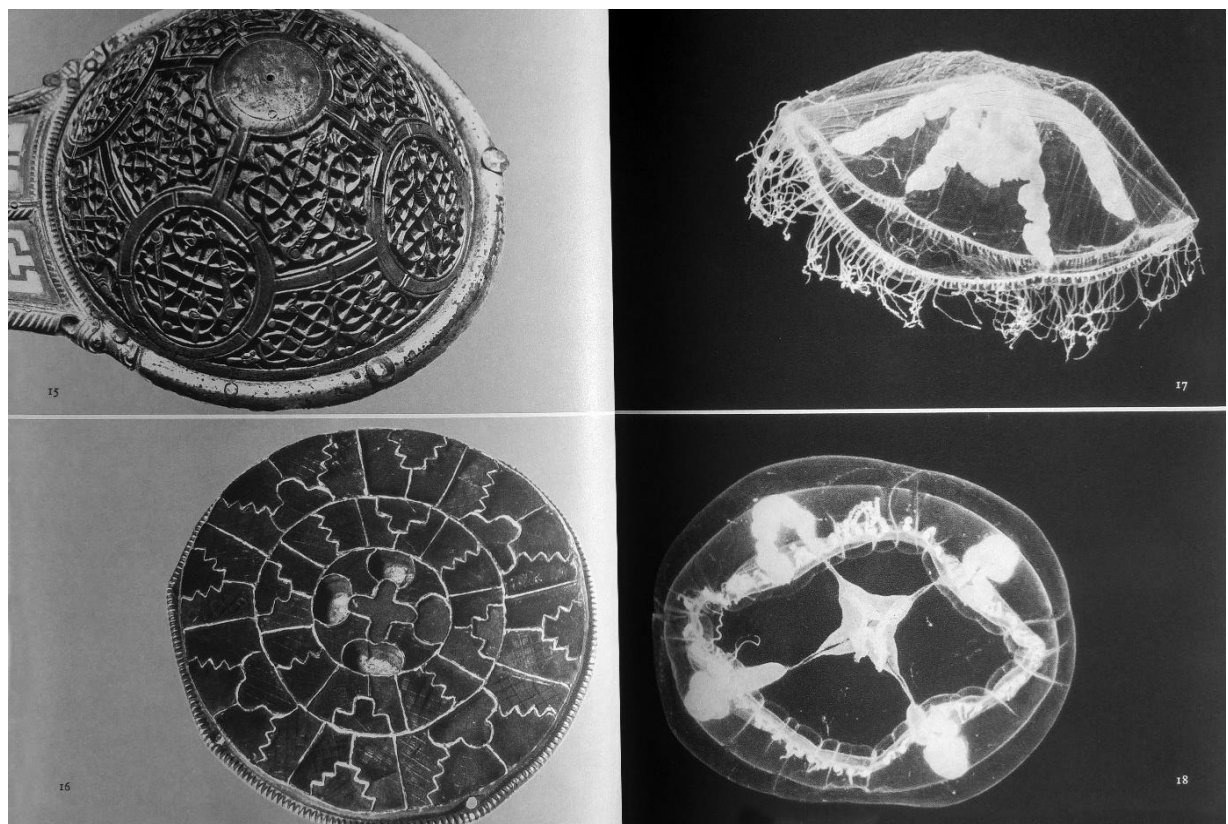

**Fig. 19 :** Zodiaque, *Introduction au monde des symboles* (G. de Champeaux, 1966).

---
[108] G. de CHAMPEAUX, *Introduction au monde des symboles*, La Pierre-Qui-Vire, 1966, p. 5-6.

Cette comparaison implicite était audacieuse, car tout en étant fondée sur une approche formelle, elle donnait à penser sur la profondeur de la vision cosmologique médiévale. A bien y réfléchir, il semble en effet que les images scientifiques à vocation heuristique fonctionnent, à un certain niveau, de façon analogue aux « images » médiévales. Dans les deux cas, il ne s'agit pas de seulement de témoigner, mais de montrer une « réalité » (dans le cas contemporain), ou plutôt une « vérité » (dans le cas médiéval), qui se joue au-delà du visuel, mais qui reste inatteignable sans son truchement. Certes, dans un cas il s'agit du Divin, dans l'autre d'un objet que le scientifique s'est choisi. Mais il me semble que le processus est relativement analogue : une matérialisation de l'invisible[109].

Or, faire apparaître l'invisible est un programme extrêmement contemporain, qui correspond à l'ère des données telle que nous la vivons aujourd'hui. Des entreprises comme Google, Facebook ou Twitter vivent de ce paradigme issu de la seconde guerre mondiale et des travaux de Claude Shannon [1916-2001] et d'Alan Turing [1912-1954], en vendant des modèles abstraits, issus de la fouille de données. Cependant, cette ère des données plonge nous semble-t-il ses racines dans les découvertes des années 1880-1930. Faire apparaître les structures invisibles, grâce à des figures visuelles, n'est-ce pas déjà le programme de la psychologie de la forme ou encore de la linguistique Sausurrienne ?[110] Cette autre manière de penser les visuels, très différente de l'approche classique, qui ne sont alors plus seulement le reflet d'un moment intellectuel, mais reflète le moment intellectuel lui-même, pourrait jouer un rôle de plus en plus important dans les décennies à venir, au-delà des sciences des données.

---

[109] H.L. KESSLER, *Seeing Medieval Art*, Peterborough, 2004 ; *ID.*, *Spiritual Seeing: Picturing God's Invisibility in Medieval Art*, Philadelphia, 2000 ; E. de BRUYNE, *Études d'esthétique médiévale*, 2 volumes, Paris, 1998 (première édition à Bruges, 1946) ; D. MEHU, « L'évidement de l'image ou la figuration de l'invisible corps du Christ (IXᵉ-XIᵉ siècle) », in *Images* re-*vues, histoire, anthropologie et théorie de l'art*, vol. 11, 2013 (en ligne : https://imagesrevues.revues.org/3384).

[110] F. de SAUSSURE, *Cours de linguistique générale. Édition critique préparée par Tullio de Mauro*, Paris, 1972.